\documentclass[a4paper,11pt]{article}
\usepackage{jheppub}
\usepackage[T1]{fontenc}
\usepackage{booktabs}

\def\CF{\mathrm{C_F}}
\def\CFsq{\mathrm{C_F^2}}
\def\CA{\mathrm{C_A}}
\def\CAsq{\mathrm{C_A^2}}
\def\Nc{\mathrm{N_c}}
\renewcommand{\d}{\mathrm{d}}

\def\A{\mathcal{A}}

\def\B{\mathcal{B}}
\def\C{\mathcal{C}}

\def\cC{\mathcal{C}}
\def\inn{\mathrm{in}}
\def\out{\mathrm{out}}

\def\as{\alpha_s}

\def\cO{\mathcal{O}}

\def\R{{\scriptscriptstyle\mathrm{R}}}

\def\G{\mathcal{G}}

\title{Azimuthal decorrelation between a jet and a $\boldsymbol Z$ boson at hadron colliders}

\author[a]{Hamza Bouaziz,}
\author[a,1]{Yazid Delenda\note{Corresponding author.}}
\author[b]{and Kamel Khelifa-Kerfa}

\affiliation[a]{Laboratoire de Physique des Rayonnements et de leurs Int\'{e}ractions avec la Mati\`{e}re\\
D\'{e}partement de Physique, Facult\'{e} des Sciences de la Mati\`{e}re\\
Universit\'{e} de Batna-1, Batna 05000, Algeria}
\affiliation[b]{D\'{e}partement de Physique, Facult\'{e} des Sciences et Technologies\\
Universit\'{e} Ahmed Zabana de Relizane, Relizane 48000, Algeria}

\emailAdd{hamza.bouaziz@univ-batna.dz}
\emailAdd{yazid.delenda@univ-batna.dz}
\emailAdd{kamel.khelifakerfa@univ-relizane.dz}

\abstract{We revisit the azimuthal decorrelation $\delta\phi$ between a jet and a $Z$ boson produced at hadron colliders. Employing different recombination schemes for the jets leads to significantly different NLL-resummed predictions for the distribution of this quantity. Specifically when the jets are reconstructed with the $E$-scheme (i.e., four-momentum addition) in the $k_t$ or anti-$k_t$ clustering algorithms, then the resummation becomes highly non-trivial due to the presence of non-global and/or clustering logarithms. We evaluate these logarithms analytically at two loops and numerically to all orders in the large-$\Nc$ limit, and present a full NLL resummation of $\delta \phi$. We extend the accuracy of the perturbative expansion of the resummed distribution at fixed order to NNLL accuracy by including $\mathcal{O}(\alpha_s)$ NLO corrections obtained with \texttt{MadGraph5\_aMC@NLO}. We compare our findings with results of various Monte Carlo event generators and with experimental data from the CMS collaboration.}


\begin{document}
\maketitle
\flushbottom

\section{Introduction}

The process of production of a $Z$ boson in association with a high-$p_t$ jet in proton-proton collisions is of great importance for LHC physics. It has been used in various ways to test the Standard Model and make precision measurements. It is also a significant background to many important processes for LHC physics such as top quark and Higgs boson production. Precise phenomenological calculations as well as experimental measurements of observables related to this process can therefore be very valuable in the identification of signal events from larger background ones.

In the transverse plane to the beam, the jet and $Z$ boson are produced back-to-back at the partonic Born order with an azimuthal angle between them equal to $\pi$. A deviation from this back-to-back configuration due to additional radiation from the incoming or outgoing partons causes a decorrelation $\delta\phi$ of the azimuthal angle between the $Z$ boson and the jet. The quantity $\delta\phi$ has been studied extensively (both phenomenologically and experimentally) in various processes and between different particles and/or jets. It has been considered between di-jets produced in deep-inelastic $e-p$ scattering (DIS) \cite{Aktas:2003ja,Banfi:2008qs} as a probe of small-Bjorken-$x$ BFKL dynamics, and in hadron-hadron \cite{Abazov:2004hm,ATLAS:2011kzm,CMS:2016adr,CMS:2016qng,ATLAS:2018sjf, Sirunyan:2019rpc,ATLAS:2019jgo} and ion-hadron \cite{ATLAS:2019jgo} collisions. Using this quantity the ATLAS collaboration measured the strong coupling $\alpha_s$ and its running at high energy scales \cite{ATLAS:2018sjf}. Azimuthal decorrelation between hadrons (instead of jets) has also been studied in di-jet production in DIS and in $e^+e^-$ annihilation \cite{Banfi:2002vw,AMY:1995trs}. Additionally, lepton-jet azimuthal decorrelation has been studied in DIS single-jet production \cite{Liu:2020dct,H1:2021wkz}. The observable of interest in this paper, i.e., the azimuthal decorrelation between a $Z$ boson and a jet produced in hadron-hadron collisions, has been studied in refs. \cite{ATLAS:2011qcw,Chatrchyan:2013tna,Sun:2018icb,Chien:2020hzh,Yang:2022qgk,Chien:2022wiq}.

In this paper we are specifically interested in the resummation of the large logarithms $L=\ln(1/\delta\phi)$ in the distribution of the azimuthal decorrelation between the $Z$ boson and the leading-$p_t$ jet, taking the form $\alpha_s^nL^m$, with $m\leq 2\,n$ at each perturbative order $n$. Being sensitive to both soft and collinear emissions to the hard partons in the event, the $\delta\phi$ distribution receives up to double logarithmic contributions. A next-to-leading logarithmic (NLL) resummation ensures that all the leading double logarithms $\alpha_s^n L^{n+1}$ and next-to-leading single logarithms $\alpha_s^n L^n$ are fully accounted for in the \emph{exponent} of the distribution. The said logarithms are enhanced near the back-to-back threshold limit where $\delta\phi$ is small. In this region there are actually two different mechanisms that produce a small $\delta\phi$: (a) suppression of strongly-ordered soft gluon emissions, and (b) vectorial cancellation of two or more hard emissions.

When the jets are reconstructed  with the $p_t$-weighted scheme in sequential recombination algorithms such as the $k_t$ \cite{Catani:1993hr,Ellis:1993tq} or anti-$k_t$ \cite{Cacciari:2008gp} clustering algorithms, the resummation of the azimuthal decorrelation is relatively straightforward. For instance, in ref. \cite{Banfi:2008qs} the di-jet azimuthal decorrelation in DIS was fully resummed at NLL accuracy. Additionally, the resummation of the $Z$-jet azimuthal decorrelation has been achieved in the transverse-momentum-dependent factorisation formalism \cite{Sun:2018icb}, and in the soft-collinear effective theory formalism up to NNLL accuracy \cite{Chien:2020hzh,Chien:2022wiq}, where in the latter the axis of the jet is defined by the direction of the highest-$p_t$ particle in the jet (known as the winner-take-all recombination scheme \cite{Bertolini:2013iqa}).

Complications in the resummation of the said quantity appear when considering other more commonly used jet recombination schemes, specifically the $E$-scheme, in which the four-momentum of the jet is defined to be the sum of the four-momenta of its constituent particles. Here, as we show in the text, the azimuthal decorrelation becomes sensitive only to gluon emissions that do not get clustered to the outgoing hard jet. Consequently, the azimuthal decorrelation in this case becomes a non-global observable and its resummation requires the treatment of non-global logarithms (NGLs) \cite{Dasgupta:2001sh,Dasgupta:2002bw} and clustering logarithms (CLs) \cite{Banfi:2005gj,Delenda:2006nf}. Such a resummation is in general non-trivial and is usually carried out numerically in the large-$\Nc$ approximation ($\Nc$ is the number of quark colours). Achieving finite-$\Nc$ numerical resummation of NGLs has been possible both in $e^+e^-$ annihilation \cite{Hatta:2013iba,Hagiwara:2015bia} and hadron-hadron collisions \cite{Hatta:2020wre}, for few observables. Furthermore, numerical resummation of sub-leading NGLs (at large $\Nc$) was achieved in refs. \cite{Banfi:2021owj,Banfi:2021xzn}. Additionally, in the context of boson-jet correlations at hadron colliders, the resummation of NGLs has been discussed in ref. \cite{Chien:2019gyf} for the transverse momentum and azimuthal decorrelation of the boson + jet system.

In this paper we consider the azimuthal decorrelation distribution when the jets are clustered with the $E$-scheme in the $k_t$ or anti-$k_t$ algorithms. We perform an NLL resummation of this distribution in the Fourier space conjugate to $\delta\phi$. We compute at two-loops the coefficients of NGLs, in the $k_t$ and anti-$k_t$ algorithms, and CLs, in the $k_t$ algorithm, as a function of the jet radius $R$. We use a numerical Monte Carlo (MC) code \cite{Dasgupta:2001sh} in order to estimate the all-orders resummed NGLs. We also improve the accuracy of the fixed-order expansion of our distribution up to NNLL accuracy, i.e., controlling up to $\alpha_s^n L^{2n-2}$ at all orders.  We additionally verify the validity of our resummation by comparing its one-loop expansion with a fixed-order MC distribution obtained with \texttt{MadGraph5\_aMC@NLO} \cite{Alwall:2014hca} and \texttt{MCFM} \cite{Campbell:2019dru}. This confirms that our resummation correctly captures the double and single logarithms, at least at one loop.  We perform the convolution of the resummed result, including NGLs and CLs, with the Born cross-section, and compare our NLL-resummed distribution with the results obtained with \texttt{Pythia 8} \cite{Sjostrand:2014zea,Alwall:2008qv} and \texttt{Herwig++} \cite{Bahr:2008pv,Bellm:2015jjp} parton showers interfaced with \texttt{MadGraph5\_aMC@NLO} \cite{Alwall:2014hca} at next-to-leading order (NLO), and with \texttt{Sherpa} stand-alone parton shower \cite{Sherpa:2019gpd}. Our resummed result shows a good agreement with the MC parton-shower distributions for a wide range of the observable. We also provide a comparison of our resummed result with the experimental data from the CMS collaboration for the same distribution \cite{Chatrchyan:2013tna}.

This paper is organised as follows. In the next section we give the precise definition of the azimuthal decorrelation between the $Z$ boson and the jet and show how this definition changes over different recombination schemes of the jet algorithm, and the impact of this on the resummation. In section 3 we perform the resummation of the large global logarithms up to NLL accuracy in the Fourier space of the azimuthal decorrelation variable. In section 4 we evaluate at two loops the NGLs and CLs as a function of the jet radius, and discuss the all-orders treatment of these logarithms. In section 5 we invert the resummed result from Fourier space back to $\delta\phi$ space. We also include the fixed-order NLO corrections in our distribution and show that our resummation correctly reproduces all the large logarithms at $\mathcal{O}(\alpha_s)$ by comparing to the NLO distribution of $\delta\phi$ obtained with  \texttt{MadGraph5\_aMC@NLO} and \texttt{MCFM}. In section 6 we compare our resummed distribution with results from various MC event generators and with experimental data from the CMS collaboration. Finally, in section 7, we draw our conclusions.

\section{Kinematics and observable definition}

\begin{figure}[ht]
\centering
\includegraphics[width=0.45\textwidth]{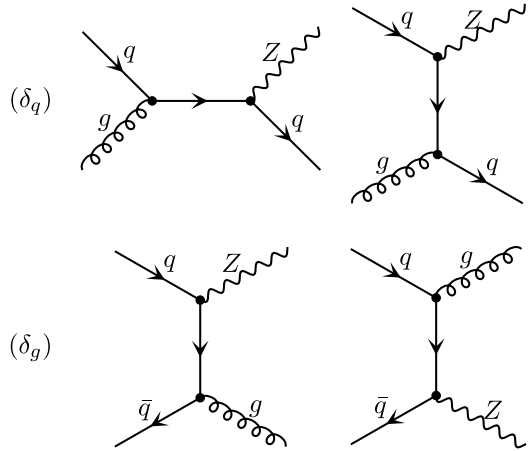}
\caption{\label{fig:01} Feynman diagrams contributing to the partonic Born process for $Z$+jet production at hadron colliders.}
\end{figure}

In this paper we consider the process of production of a $Z$ boson and a jet $J$ at hadron colliders. There are two partonic channels contributing to the Born process, as shown in figure \ref{fig:01}, specifically $(\delta_g):q\bar{q}\to Zg$ and $(\delta_q):qg\to Zq$. \footnote{The channel $(\delta_q)$ also includes incoming anti-quarks, i.e. $\bar{q}g\to Z\bar{q}$.} Beyond the Born level this process is showered by soft/collinear emissions from both initial-state and final-state hard partons. For the purpose of this paper, and in order to achieve NLL accuracy, it suffices to assume the eikonal approximation in which the emissions are strongly ordered such that at order $n$ we have $k_{tn}\ll\cdots\ll k_{t2}\ll k_{t1}\ll p_t$. The transverse momenta of the outgoing particles in the plane perpendicular to the beam axis are given by
\begin{subequations}
\begin{align}
\vec{\mathbf{p}}_{tZ}&=\tilde{p}_{t}\left(1,0\right),\\
\vec{\mathbf{p}}_{tJ}&=p_{t}\left(\cos(\pi-\epsilon),\sin(\pi-\epsilon)\right),\\
\vec{\mathbf{k}}_{ti}&=k_{ti}\left(\cos\phi_i,\sin\phi_i\right),
\end{align}
\end{subequations}
with $\tilde{p}_{t}$, $p_t$, and $k_{ti}$ being, respectively, the transverse momenta of the $Z$ boson, the hard parton initiating the jet $J$, and the emitted soft gluon $(i)$. Moreover, $\phi_i$ is the azimuthal angle of the soft emission $(i)$ measured with respect to the beam axis. At the Born level the $J$ and $Z$ are exactly back-to-back in the transverse plane, that is, the azimuthal angle between them is exactly $\pi$. At higher orders, when there are accompanying soft emissions, a small ``decorrelation'' $|\epsilon|\ll 1$ between them occurs. Without loss of generality we fixed the azimuthal angle of the $Z$ boson to be $0$ and that of the hard parton $J$ to be $\pi-\epsilon$. From conservation of momentum, specifically the $y$ component, we infer that $\epsilon$ is expressed at order $n$ in terms of the transverse momenta $k_{ti}$ at NLL accuracy as
\begin{equation}
\epsilon=-\sum_i^n\frac{k_{ti}}{p_t}\,\sin\phi_i\,.
\end{equation}

When clustering the jets with a given jet algorithm, there are three commonly-used recombination schemes. In the $E$-scheme, which is the most common one, the four-momentum of the jet is simply defined to be the sum of the four-momenta of its particle constituents, $p_J=\sum_{i\in J} p_i$. The jet transverse momentum $p_{tJ}$, rapidity $\eta_J$, and azimuthal angle $\phi_J$ are then deduced from the resulting four-momentum. In the $p_t$-weighted scheme, the $p_{tJ}$, $\eta_J$, and $\phi_J$ of the jet are defined by
\begin{subequations}
\begin{align}
p_{tJ}&=\sum_{i\in J}p_{ti}\,,\label{eq:pts}\\
\eta_J&=\frac{1}{p_{tJ}}\sum_{i\in J}{p_{ti}\,\eta_i}\,,\\
\phi_J&=\frac{1}{p_{tJ}}\sum_{i\in J}{p_{ti}\,\phi_i}\,.
\end{align}
\end{subequations}
Additionally there is the ``Winner-Take-All'' (WTA) scheme \cite{Bertolini:2013iqa} in which the direction of the jet axis (and thus the jet rapidity and azimuth) is defined to be that of the hardest parton within it, and its transverse momentum is the scalar sum of its constituent transverse momenta as in eq. \eqref{eq:pts}. \footnote{In all cases, the recombination scheme is applied at each intermediate iteration step to define pseudo-jets during the clustering.} The ATLAS and CMS collaborations at the LHC and the D\O\, collaboration at the Tevatron employ the $E$-scheme, while the H1 collaboration at HERA uses the $p_t$-weighted scheme.

In the $p_t$-weighted scheme, the $Z$-jet azimuthal decorrelation observable $\delta\phi$ may be shown to depend on the emission of the soft gluons at order $n$ as
\begin{equation}
\delta\phi^{p_t\text{-weighted}}=\left|\sum_i^n\frac{k_{ti}}{p_t}\left(\sin\phi_i-\Theta_{iJ}\,(\pi-\phi_i)\right)\right|.
\end{equation}
Furthermore, in the WTA scheme, the observable becomes
\begin{equation}
\delta\phi^{\mathrm{WTA}}=\left|\sum_i^n\frac{k_{ti}}{p_t}\sin\phi_i\right|.
\end{equation}
In the above two definitions, the sum over $i$ extends over \emph{all} emissions in the event, and $\Theta_{iJ}=1$ if gluon $i$ is clustered into the jet $J$ and $0$ otherwise. Being inclusive over emissions in the entire angular phase space, the definitions above make the observable \emph{continuously} global, \footnote{In the $p_t$-weighted scheme, although the dependence of $\delta\phi$ on emissions inside and outside the jet are different, the observable is still global because its dependence on $k_t$ is linear in both cases.} as has been established in the literature \cite{Banfi:2008qs,Chien:2020hzh,Chien:2022wiq,Banfi:2004yd}. Consequently, the resummation of this observable in these two cases is relatively straightforward since it does not require the non-trivial treatment of NGLs and/or CLs. We leave this study for our future work.

Employing the $E$-scheme, on the other hand, the decorrelation is caused only by the emitted soft gluons that do not get clustered to the jet. We thus define the observable to be studied in this paper as
\begin{equation}\label{eq:addef}
\delta\phi=\left|\sum_{i\notin J}^n\frac{k_{ti}}{p_t}\,\sin\phi_i\right|,
\end{equation}
where the sum excludes all emissions that end up inside the jet after applying the jet algorithm. This situation is similar to gap-energy observables which are sensitive to emissions in the rapidity gap between two jets. Therefore the observable under consideration is non-global \cite{Dasgupta:2001sh,Dasgupta:2002bw}, and to achieve NLL accuracy one is forced to properly address the resummation of NGLs and/or CLs. Our aim in this paper is to treat this very issue.

\section{Global resummation}

At small values of $\delta\phi$ one may expect the distribution to behave like a Sudakov form factor. While this is correct for a wide range of values of $\delta\phi$ this observation actually fails at very small $\delta\phi$. To explain this, we note that the azimuthal decorrelation $\delta\phi$ as defined in eq. \eqref{eq:addef} is the \emph{algebraic} sum of the $y$ components of momenta of emitted gluons, normalised to $p_t$. This implies that the small values of $\delta\phi$ may well be obtained by cancellation of hard emissions instead of suppression of soft emissions. This observation is similar to that made by Parisi and Petronzio in ref. \cite{Parisi:1979se} for transverse momentum distributions. The implication of this fact is the failure of the strong hierarchy of the logarithmic accuracy, $\mathrm{LL}\gg \mathrm{NLL}\gg\mathrm{NNLL}\cdots$, in the distribution at very small values of $\delta\phi$. Nevertheless, as we shall show, at small and intermediate values of $\delta\phi$ the resummation based on the soft gluon suppression, and combined with fixed-order NLO effects, provides a good description of the distribution obtained with MC parton showers.

To achieve NLL resummation we need to treat three types of gluon emissions off the three primary hard partons, namely
\begin{itemize}
\item primary soft and/or collinear emissions outside the jet, resulting in global logarithms,
\item primary soft wide-angle emissions inside or outside the jet, resulting in CLs,
\item secondary non-Abelian soft emissions, resulting in NGLs.
\end{itemize}

In this section we address the former contribution and leave the discussion of the other two contributions to the next section. To compute the differential cross-section $1/\sigma\,\d\sigma/\d\delta\phi$ we first evaluate the integrated distribution defined by the cross-section $\sigma(\Delta)$ for events with azimuthal decorrelation $\delta\phi$ being less than some value $\Delta$. To this end we compute the integrated cross-section (following the notation adopted in refs. \cite{Banfi:2004yd,Ziani:2021dxr,Dasgupta:2012hg})
\begin{equation}\label{eq:master}
\sigma_{\mathrm{sc}}^{\R}(\Delta)=\sum_{\delta}\int\d\B_{\delta}\,\frac{\d\sigma_{0\delta}}{\d\B_{\delta}}\,\Xi_\B\sum_{n=1}^\infty\int\d P_n\,\Theta\left(\Delta-\left|\sum_{i\notin J}^n \frac{k_{ti}}{p_t}\, \sin\phi_i\right|\right),
\end{equation}
where the subscript ``sc'' indicates that we have only considered soft-collinear emissions, and the superscript ``R'' indicates real-emission contributions. Virtual corrections and hard collinear contributions will be included later. Here $\d\sigma_{0\delta}/\d\B_{\delta}$ is the differential Born cross-section for the process channel $\delta$ and $\Xi_\B$ denotes kinematical cuts. The explicit expression for $\d\sigma_{0\delta}/\d\B_{\delta}$ together with the differential Born configuration $\d\mathcal{B}_\delta$ are presented in detail in ref. \cite{Ziani:2021dxr}. In the eikonal approximation, one can write the probability of independent emission of $n$ primary soft gluons off the three hard partons ($q,\tilde{q},g$) outside the jet as \cite{Khelifa-Kerfa:2020nlc,Banfi:2000si}
\begin{equation}
\d P_n=\frac{1}{n!}\prod_{i=1}^{n}\,\sum_{(\alpha\beta)}\mathcal{C}_{\alpha\beta}\,\frac{\as(\kappa_{ti,\alpha\beta}^2)}{\pi}\,\frac{\d k_{ti}}{k_{ti}}\,\d\eta_i\,\frac{\d\phi_i}{2\pi}\,w_{\alpha\beta}^i\,\Theta_{\out}(k_i)\,,
\end{equation}
where $\sum_{(\alpha\beta)}$ is the sum over the three dipoles formed by the three hard legs ($q,\tilde{q},g$), and $\mathcal{C}_{\alpha\beta}$ is the colour factor associated with the dipole $(\alpha\beta)$. We have $\mathcal{C}_{q\tilde{q}}=-1/\Nc$ for dipoles involving quarks only and $\mathcal{C}_{qg}=\Nc=3$ for dipoles involving a quark and a gluon. Here $\as$ is the strong coupling defined in the bremsstrahlung Catani-Marchesini-Webber (CMW) scheme \cite{Catani:1990rr}, and its argument is the invariant transverse momentum of the emission $(i)$ with respect to the dipole emitting it \cite{Catani:1989ne,Catani:1992ua,Catani:1999ss}, $\kappa_{ti,\alpha\beta}^2=k_{ti}^2/\omega_{\alpha\beta}^i$, where the antenna function is defined by
\begin{equation}
w_{\alpha\beta}^i=\frac{k_{ti}^2}{2}\,\frac{p_\alpha\cdot p_\beta}{(p_\alpha\cdot k_i)(p_\beta\cdot k_i)}\,.
\end{equation}
We restrict all emissions to be outside the hard jet with the step function
\begin{equation}
\Theta_{\out}(k_i)=\Theta\left[(\eta_i-y)^2+(\phi_i-\pi)^2-R^2\right],
\end{equation}
with $R$ the jet radius and $\eta_i$ and $y$ the rapidities of gluon $(i)$ and the outgoing hard parton initiating the jet, respectively. While this correctly captures all logarithms originating from primary emissions in the anti-$k_t$ clustering algorithm, it still needs corrections from CLs contributing at the NLL level when employing the $k_t$ algorithm, as we shall explain in the next section.

To proceed we factorise the step function using its Fourier representation by writing
\begin{equation}
\Theta\left(\Delta-\left|\sum_{i\notin J}^n\frac{k_{ti}}{p_t}\,\sin\phi_i\right|\right)=\int_{-\infty}^\infty \frac{\d b}{\pi\,b}\sin(b\,\Delta)\prod_{i\notin J}^n e^{i\,b\,k_{ti}\,\sin\phi_i/p_t}\,.
\end{equation}
This factorised form, together with the factorised emission amplitude squared and phase space, allows us to exponentiate the integrated distribution as follows
\begin{equation}\label{eq:master2}
\sigma_{\mathrm{sc}}(\Delta)=\sum_{\delta}\int\d\B_{\delta}\,\frac{\d\sigma_{0\delta}}{\d\B_{\delta}}\,\Xi_\B\int_{-\infty}^\infty\frac{\d b}{\pi\,b}\sin(b\,\Delta)\exp\left[-\mathcal{R}_\delta(b)\right],
\end{equation}
where the radiator for channel $\delta$ is given by
\begin{equation}\label{eq:bradiator}
\mathcal{R}_\delta(b)=-\sum_{(\alpha\beta)}\mathcal{C}_{\alpha\beta}\int\frac{\as(\kappa_{t,\alpha\beta}^2)}{\pi}\,\frac{\d k_t}{k_t}\,\d\eta\,\frac{\d\phi}{2\pi}\,\Theta_{\out}(k)\,w_{\alpha\beta}^k \left(e^{i\,b\,k_t\,\sin\phi/p_t}-1\right).
\end{equation}
The integration over $k_t$ extends from $0$ to the hard scale $p_t$. Notice that our resummed formula \eqref{eq:master2} now includes virtual corrections at all orders through the last term $(-1)$ that appears in the radiator \eqref{eq:bradiator}.

The next step is to employ the following approximation, which is valid at NLL accuracy,
\begin{equation}
-\left(e^{i\,b\,k_{t}\,\sin\phi/p_t}-1\right)\approx \Theta\left(k_{t}\,|\sin\phi|-p_t/\bar{b}\right),
\end{equation}
where $\bar{b}=b\,e^{\gamma_E}$ and $\gamma_E\approx0.577$ is the Euler-Mascheroni constant. This enables us to reduce the range of integration over $b$ by making the replacement $\int_{-\infty}^\infty \d b \to 2\int_{0}^\infty \d b$ since the integrand becomes an even function in $b$.

In appendix \ref{sec:rad} we perform at NLL accuracy the integrations over $k_t$, $\eta$ and $\phi$ for the three dipole contributions to the radiator, and present its expression in $b$ space in the standard form. Note that including \emph{hard-collinear} emissions (as we do in the appendix) involves evolving the scale of the parton distribution functions (PDFs) from the factorisation scale $\mu_{\mathrm{f}}$ to $\mu_{\mathrm{f}}/\bar{b}$ \cite{Banfi:2001ci}. Further details about this and about how to perform the $b$ integration will be discussed in section 5. In the next section, we consider corrections to the radiator due to CLs and NGLs at two loops and to all orders.

\section{Non-global and clustering logarithms}

\subsection{Non-global logarithms at two loops}

At $\mathcal{O}(\alpha_s^2)$, NGLs \cite{Dasgupta:2001sh,Dasgupta:2002bw} originate when a soft gluon $k_1$ is emitted inside the jet, which itself coherently emits another softer gluon $k_2$ outside of it without being clustered back to the jet \cite{Appleby:2002ke}. When the softer emission is real it causes a small decorrelation of the azimuthal angle between the jet and the $Z$ boson, and when it is virtual the decorrelation is zero, leading to a real-virtual mismatch in the contribution to the integrated distribution. This configuration is opposite to the jet mass observable where NGLs originate when the harder emission is outside the jet while the softer is inside \cite{Dasgupta:2012hg, Ziani:2021dxr}.

The non-global logarithmic contribution to the integrated distribution at $\mathcal{O}(\alpha_s^2)$ may be written as follows \cite{Dasgupta:2012hg}
\begin{equation}
\frac{1}{\sigma_{0,\delta}}\,\sigma_{2,\delta}^{\mathrm{NG}}(\Delta)=-\frac{1}{2!}\,\frac{\alpha_s^2}{\pi^2}\,\ln^2\frac{1}{\Delta}\,\mathcal{G}_2^\delta(R)\,,
\end{equation}
where $\sigma_{0,\delta}$ denotes the Born cross-section for channel $\delta$. In the anti-$k_t$ algorithm the radius-dependent function $\mathcal{G}_2^\delta(R)$ is expressed as
\begin{align}\label{eq:G2_akt}
\mathcal{G}^{\delta,\mathrm{ak_t}}_2(R)&=\CA\sum_{(ij)}\cC_{ij}\,R^4\int_0^1 r_1\,\d r_1\,\frac{\d\theta_1}{2\pi}\,\int_1^{\pi/\left(R\,|\sin\theta_2|\right)}r_2\,\d r_2\,\frac{\d\theta_2}{2\pi}\,\A_{ij}^{12}\notag\\
&\equiv\CA\sum_{(ij)}\cC_{ij}\,\mathcal{G}_{2}^{(ij),\mathrm{ak_t}}(R)\,,
\end{align}
where $\CA=\Nc$ is the colour factor for the secondary gluon emission, $r_i$ and $\theta_i$ are polar variables in the $(\eta_i-\phi_i)$ plane defined by $\eta_i=y+R\,r_i\cos\theta_i$ and $\phi_i=\pi+R\,r_i\sin\theta_i$, and $\A_{ij}^{12}$ is the irreducible antenna function for the correlated emission of the two gluons $k_1$ and $k_2$ off the hard dipole $(ij)$
\begin{equation}
\A_{ij}^{12}=w^1_{ij}\left(w^2_{i1}+w^2_{1j}-w^2_{ij}\right).
\end{equation}
Notice that $\A_{ij}^{12}$ is symmetric under the exchange $k_1\leftrightarrow k_2$. This means that the integral \eqref{eq:G2_akt} is exactly identical to that for the jet mass observable in refs. \cite{Dasgupta:2012hg, Ziani:2021dxr}, with the change of variable $k_1\leftrightarrow k_2$. Hence, at two loops, the coefficient of NGLs in the anti-$k_t$ algorithm is identical for both observables. We expect, however, NGLs to be different at higher orders since the phase space is not symmetric under the exchange of the gluons and, at finite $\Nc$, the squared amplitudes too are not symmetric.

The expressions for the dipole functions $\mathcal{G}_{2}^{(ij)}(R)$ have been computed in refs. \cite{Dasgupta:2012hg,Ziani:2021dxr} and the reported results are as follows
\begin{subequations}
\begin{align}
\mathcal{G}^{(ab),\mathrm{ak_t}}_2&=-R^2\ln R+0.500\,R^2+0.125\,R^4-0.003\,R^6+\mathcal{O}(R^8)\,,\\
\mathcal{G}^{(aj),\mathrm{ak_t}}_2&=\mathcal{G}^{(bj),\mathrm{ak_t}}_2=0.822+0.003\,R^4+\cO(R^8)\,.
\end{align}
\end{subequations}
Combining the results for the different channels one has
\begin{subequations}\label{eq:NGLs_akt_Channels}
\begin{align}
\mathcal{G}^{\delta_g,\mathrm{ak_t}}_2&=\CF\,\CA\left[-2\,R^2\ln R+R^2+0.250\,R^4-0.007\,R^6+\mathcal{O}(R^8)\right]+\notag\\
&+\CAsq\left[1.645+R^2\ln R-0.500\,R^2-0.118\,R^4+0.003\,R^6+\mathcal{O}(R^8)\right],\\
\mathcal{G}^{\delta_q,\mathrm{ak_t}}_2&=\CF\,\CA\left[1.645+0.007\,R^4+\mathcal{O}(R^8)\right]+\notag\\
&+\CAsq\left[-R^2\ln R+0.500\,R^2+0.125\,R^4-0.003\,R^6+\mathcal{O}(R^8)\right],
\end{align}
\end{subequations}
where $\CF = (\mathrm{N_c^2}-1)/(2\,\Nc)$.

In the $k_t$ algorithm, the clustering of particles starts with the softest emissions. Thus the coefficient of NGLs at two loops reads
\begin{align}\label{eq:G2_kt}
\mathcal{G}^{\delta,\mathrm{k_t}}_2(R)&=\CA\sum_{(ij)}\cC_{ij}\,R^4\int_0^1 r_1\,\d r_1\,\frac{\d\theta_1}{2\pi}\int_1^{\pi/\left(R\,|\sin\theta_2|\right)}r_2\,\d r_2\,\frac{\d\theta_2}{2\pi}\,\Theta\left(d_{12}-d_2\right)\A_{ij}^{12}\notag\\
&=\CA\sum_{(ij)}\cC_{ij}\,\mathcal{G}_{2}^{(ij),\mathrm{k_t}}(R)\,,
\end{align}
where the distance measures are defined by $d_i=k_{ti}^2\,R^2$ and $d_{ij}=\mathrm{min}(k_{ti}^2,k_{tj}^2)\left(\delta \eta_{ij}^2+\delta \phi_{ij}^2\right)$. The step function $\Theta\left(d_{12}-d_2\right)$ forbids the emission $k_2$ from being clustered to the harder one $k_1$, and subsequently to the jet, resulting in the large NGLs as explained above. Here, the clustering step function is not related by symmetry to that for the jet mass observable \cite{Ziani:2021dxr} (given by $\Theta(d_{12}-d_{2j})$), and thus the result is going to be different. In terms of the polar variables, this step function is given by
\begin{equation}
\Theta\left(d_{12}-d_2\right)=\Theta\left[r_1^2+r_2^2-2\,r_1\,r_2\cos(\theta_1-\theta_2)-1\right].
\end{equation}
In analogy to the anti-$k_t$ clustering case, we perform the integrations by expanding the antenna function as a power series in the jet radius. We obtain the following results for the various dipole contributions
\begin{subequations}
\begin{align}
\mathcal{G}^{(ab),\mathrm{k_t}}_2&=-R^2\,\ln R -0.128\,R^2+0.177\,R^4-0.004\,R^6+\mathcal{O}(R^8)\,,\\
\mathcal{G}^{(aj),\mathrm{k_t}}_2&=\mathcal{G}^{(bj),\mathrm{ak_t}}_2=0.183-0.121\,R^2+0.007\,R^4+0.0003\,R^6+\cO(R^8)\,.
\end{align}
\end{subequations}
In terms of channels we may write
\begin{subequations}\label{eq:NGLs_kt_Channels}
\begin{align}
\mathcal{G}^{\delta_g,\mathrm{k_t}}_2&=\CF\,\CA\left[-2\,R^2\ln R-0.255\,R^2+0.353\,R^4-0.009\,R^6+\mathcal{O}(R^8)\right]+\notag\\
&+\CAsq\left[0.366+R^2\ln R-0.115\,R^2-0.163\,R^4+0.005\,R^6+\mathcal{O}(R^8)\right],\\
\mathcal{G}^{\delta_q,\mathrm{k_t}}_2&=\CF\,\CA\left[0.366-0.243\,R^2+0.014\,R^4+0.001\,R^6+\mathcal{O}(R^8)\right]+\notag\\
&+\CAsq\left[-R^2\ln R-0.128\,R^2+0.177\,R^4-0.004\,R^6+\mathcal{O}(R^8)\right].
\end{align}
\end{subequations}

\begin{figure}[ht]
\centering
\includegraphics[width=0.55\textwidth]{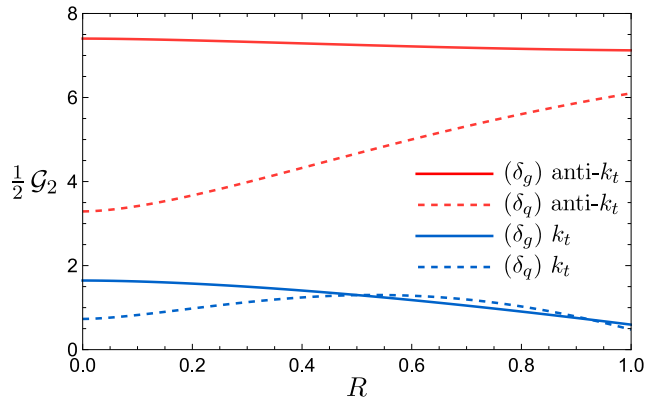}
\caption{\label{fig:02} Coefficient of NGLs at two loops for the two algorithms $k_t$ and anti-$k_t$ clustering, and for the two channels $(\delta_q)$ and $(\delta_g)$.}
\end{figure}

Note here that, as apposed to the jet mass variable, the small-$R$ limit of this result is half that reported for the jet mass distribution in $e^+e^- \to$ di-jet events. We show in figure \ref{fig:02} a plot of the overall coefficient of NGLs, $\frac{1}{2}\,\mathcal{G}_2$, as a function of the jet radius $R$ for the two algorithms and for the two channels. We observe here the sizable coefficient of NGLs in the case of anti-$k_t$ clustering compared to $k_t$ clustering. This observation has been made in previous studies of NGLs with $k_t$ clustering \cite{Appleby:2002ke,Delenda:2006nf}. The latter algorithm tends to reduce the size of NGLs while resulting in another tower of large logarithms known as CLs. This observation implies that using $k_t$ clustering is in fact phenomenologically favoured over anti-$k_t$ since the all-orders resummation of NGLs is in general less accurate than that of CLs, as one has to employ an approximation such as the large-$\Nc$ limit.

\subsection{Clustering logarithms at two loops}

CLs appear at $\mathcal{O}(\alpha_s^2)$ when two soft gluons are emitted directly from the hard partons (i.e., primary emissions) and when the $k_t$ algorithm is applied on the final-state partons. The phase space that gives rise to CLs is such that the harder emission $k_1$ is initially inside the jet, $d_{1j}<d_1$, \footnote{The distances $d_i$ and $d_{ij}$ for the $k_t$ algorithm are defined in the previous subsection.} while the softer one $k_2$ is initially (before clustering) outside, $d_{2j}>d_2$, with the distance between them $d_{12}$ being less than $d_{2}$. Applying the $k_t$ algorithm, when both emissions are real, the softer particle $k_2$ is clustered to the harder one $k_1$ and the recombined pseudo-jet is (at NLL accuracy) along the direction of $k_1$, and ultimately both partons end up inside the jet leading to a zero $Z$-jet azimuthal decorrelation. However, when the particle $k_1$ is virtual there is no particle to drag the softer real emission $k_2$ into the jet, and a non-zero decorrelation occurs. Thus we obtain a real-virtual miscancellation which translates into the following contribution to the integrated cross-section at $\mathcal{O}(\alpha_s^2)$
\begin{align}\label{eq:C2}
\frac{1}{\sigma_{0,\delta}}\,\sigma^{\mathrm{CL}}_{2,\delta}(\Delta)=\frac{1}{2!}\,\frac{{\alpha}_s^2}{\pi^2}\,\ln^2\frac{1}{\Delta}\,\mathcal{F}_2^{\delta}(R)\,,
\end{align}
where
\begin{align}
\mathcal{F}_2^{\delta}(R)&=\sum_{(ij)(\ell m)}\cC_{ij}\,\cC_{\ell m}\,R^4\int_0^1r_1\,\d r_1\,\frac{\d\theta_1}{2\pi}\int_1^2r_2\,\d r_2\,\frac{\d\theta_2}{2\pi}\,\Theta\left(d_2-d_{12}\right)w_{ij}^1\,w_{\ell\,m}^2\,,\label{eq:F2}
\end{align}
and the sum extends over all 9 dipole pairs $(ij)$ and $(\ell m)$. The coefficients of CLs for the various channels are given by
\begin{subequations}\label{eq:sss}
\begin{align}
\mathcal{F}_2^{\delta_g}(R)&=\CFsq\,0.413\,R^4+\CF\,\CA\left[1.510\,R^2-0.207\,R^4+0.007\,R^6+\mathcal{O}(R^{10})\right]+\notag\\
&+\CAsq\left[0.914-0.378\,R^2+0.043\,R^4-0.002\,R^6+\mathcal{O}(R^{8})\right],\label{eq:gh}
\end{align}
for channel $(\delta_g)$,
\begin{align}
\mathcal{F}_2^{\delta_q}(R)&=\CAsq\left[0.214\,R^2+0.141\,R^4+0.001\,R^6+\mathcal{O}(R^{10})\right]+\notag\\
&+\CF\,\CA\left[0.327\,R^2+0.027\,R^4+0.001\,R^6+\mathcal{O}(R^{10})\right]+\notag\\
&+\CFsq\left[0.914+0.592\,R^2+0.081\,R^4+0.003\,R^6+\mathcal{O}(R^8)\right],\label{eq:qh}
\end{align}
for channel $(\delta_q)$.
\end{subequations}
\begin{figure}[ht]
\centering
\includegraphics[width=0.55\textwidth]{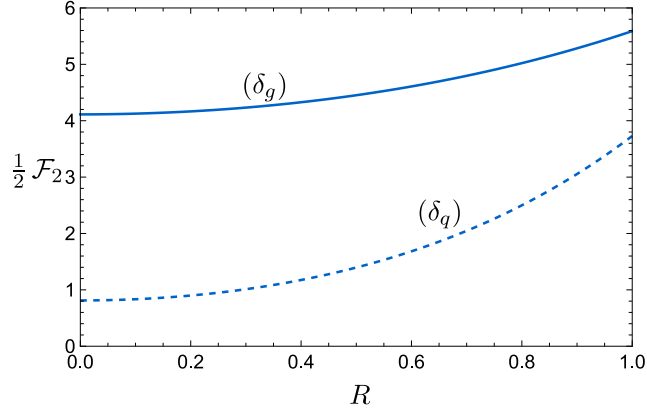}
\caption{\label{fig:03} Coefficient of CLs at two loops for the two channels $(\delta_q)$ and $(\delta_g)$.}
\end{figure}

\begin{figure}[ht]
\centering
\includegraphics[width=0.55\textwidth]{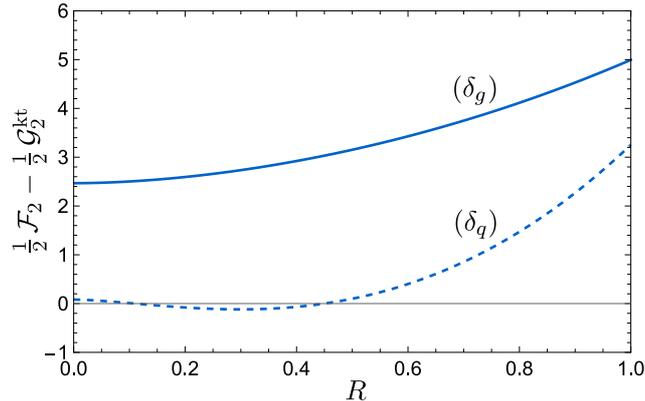}
\caption{\label{fig:04} Combined coefficient of CLs and NGLs at two loops for the two channels $(\delta_q)$ and $(\delta_g)$.}
\end{figure}

Figure \ref{fig:03} shows a plot of the coefficient of CLs at two loops as a function of the jet radius $R$ for the two channels $(\delta_q)$ and $(\delta_g)$. We observe that the gluon channel has quite a large CLs coefficient, even larger than NGLs coefficient in $k_t$ clustering by more than a factor of $2$, while the quark channel has a somewhat smaller coefficient. Both coefficients seem to be increasing with $R$. The overall coefficient of the single logarithm $\alpha_s^2/\pi^2\ln^2(1/\Delta)$ due to both NGLs and CLs, $\frac{1}{2}(\mathcal{F}_2-\mathcal{G}_2^{\mathrm{k_t}})$, is shown in figure \ref{fig:04}. For the quark channel NGLs and CLs tend to cancel each other for jet radii smaller than about $0.6$, while their combination is large for all values of $R$ in the gluon channel. This signifies the importance of the contributions of NGLs and CLs in our distribution for all values of $R$, as apposed to the jet mass observable where the combined NGLs and CLs are significant only for small values of $R$, and tend to cancel each other out at larger values of $R$ for both channels \cite{Ziani:2021dxr}.

In the next subsection we discuss the all-orders numerical resummation of NGLs and CLs in the large-$\Nc$ approximation.

\subsection{Non-global and clustering logarithms at all orders}

The calculation of NGLs and CLs beyond $\mathcal{O}(\alpha_s^2)$ is difficult due to the complexity of higher-order amplitudes of emission and phase space. NGLs originate from the coherent emission of a single soft gluon clustered outside the hard jet from a group of harder ones which are clustered into the jet. One usually uses a numerical MC approach which employs the large-$\Nc$ limit to approximate the all-orders resummation of these emissions.

Recently, a dipole-evolution program (\texttt{Gnole}) that resums NGLs at NLL and NNLL accuracy in the large-$\Nc$ limit was published \cite{Banfi:2021owj,Banfi:2021xzn}. In its current version it is only possible to perform this resummation for observables in single-dipole processes and where NGLs are leading logarithms, i.e., observables defined in regions away from all hard partons, such as rapidity-gap observables and di-jet azimuthal decorrelation in $e^+e^-$ annihilation. Future releases of this program may extend to the resummation of quantities such as the one considered here. In this work we perform the all-orders resummation of NGLs using the code upon which \texttt{Gnole} is based, and which was first developed in ref. \cite{Dasgupta:2001sh} and adapted in ref. \cite{Dasgupta:2012hg}. In the large-$\Nc$ limit, one can treat the resummation of NGLs based on the independent evolution of the three hard dipoles in the process \cite{Banfi:2002hw,Dasgupta:2012hg}.

We parameterise the numerical results for the all-orders resummed NGLs obtained for the channel $\delta$ in the anti-$k_t$ algorithm by the function \cite{Dasgupta:2001sh,Dasgupta:2012hg} \footnote{Since the program works in the large-$\Nc$ limit, and only works with quark dipoles, we set all the colour factors $\mathcal{C}_{ij}=2\,\CF=\Nc$ during the parametrisation. However, we use the actual colour factors in the resummed form factor.}
\begin{equation}
\mathcal{S}_{\delta}^{\mathrm{ak_t}}(t)=\exp\left(-\frac{t^2}{8}\,\CA\,\sum_{(ij)}\mathcal{C}_{ij}\,\G_2^{(ij),\mathrm{ak_t}}\,\frac{1+(a_{ij}\,\CA\,t/4)^2}{1+(b_{ij}\,\CA\,t/4)^{c_{ij}}}\right),
\end{equation}
where the evolution parameter is given by
\begin{equation}
t(1/\Delta) = -\frac{1}{\pi\beta_0}\ln(1-2\alpha_s\beta_0\ln(1/\Delta))\,,
\end{equation}
with $\beta_0$ the one-loop coefficient of the QCD beta function (see appendix). Fitting the numerical data to this parametrisation yields the results shown in table \ref{tab:coef} for the chosen jet radius $R=0.5$. Notice that the expansion of the parametrisation function at two loops reproduces the results obtained in the previous subsections.
\begin{table}
\centering
\begin{tabular}{ l | c | c  }
\toprule
 & $(ab)$ & $(aj)$ and $(bj)$ \\
\midrule
$a_{ij}$ & 0.446 & 0.301\\
$b_{ij}$ & 1.882 & 0.377\\
$c_{ij}$ & 1.33  & 1.33 \\
\bottomrule
\end{tabular}
\caption{Fitting parameters for the resummed NGLs for anti-$k_t$ clustering with $R=0.5$.\label{tab:coef}}
\end{table}

Performing a similar resummation in the $k_t$ algorithm is not straightforward. In order to obtain NGLs and CLs, one has to run the program to compute the overall distribution, and then divide away the global factor that is obtained by allowing only primary emissions directly off the hard Born partons. While this is straightforward in the anti-$k_t$ algorithm, the subtraction is quite difficult in \emph{multi-dipole} processes in the presence of CLs, as was mentioned in ref. \cite{Banfi:2021xzn}. We have verified that the exponential of the two-loops result of NGLs in the anti-$k_t$ algorithm does not produce noticeable differences compared to the numerical all-orders resummed result. Furthermore, the numerical all-orders resummed CLs have been shown to be well approximated by the exponential of the two-loops result in the case of jet mass distribution in di-jet production in $e^+e^-$ annihilation process \cite{Delenda:2012mm}. We thus adopt for the $k_t$ algorithm this two-loops exponential to approximate the all-orders resummed result for both CLs
\begin{equation}
\mathcal{C}_{\delta}^{\mathrm{k_t}}(t) = \exp\left(\frac{t^2}{8}\,\mathcal{F}_2^{\delta}\right),
\end{equation}
and similarly for the NGLs
\begin{equation}
\mathcal{S}_{\delta}^{\mathrm{k_t}}(t) = \exp\left(-\frac{t^2}{8}\,\mathcal{G}_2^{\delta}\right),
\end{equation}
and where we note that $\mathcal{C}_{\delta}^{\mathrm{ak_t}}(t)=1$ in the anti-$k_t$ algorithm.

\section[Result in \texorpdfstring{$\boldsymbol\Delta$}{Delta} space]{Result in $\boldsymbol\Delta$ space}

\subsection[Evaluation of the \texorpdfstring{$b$}{b} integral]{Evaluation of the $b$ integral}

One approach that can be employed to evaluate the $b$ integral analytically in order to obtain the resummed distribution in $\Delta$ space is to Taylor-expand the radiator about the saddle point $\bar{b}_0=1/\Delta$
\begin{equation}
\mathcal{R}_{\delta}(\bar{b})=\mathcal{R}_{\delta}(1/\Delta)+\mathcal{R}_{\delta}'(1/\Delta)\left(\ln\bar{b}-\ln\bar{b}_0\right),
\end{equation}
with
\begin{equation}
\mathcal{R}_{\delta}'(\bar{b})=\frac{\partial \mathcal{R}_{\delta}(\bar{b})}{\partial \ln \bar{b}}\,.
\end{equation}
In the non-global and clustering logarithmic functions as well as in PDFs one simply makes the substitution $\bar{b} \to 1/\Delta$. One can then perform the integration over $b$ in eq. \eqref{eq:master2} using
\begin{equation}
\int_0^\infty\frac{\d b}{b}\sin(b\,\Delta)\exp\left[-\mathcal{R}'_\delta\ln\bar{b}\right] =\exp\left[\left(-\gamma_E+\ln(1/\Delta)\right)\mathcal{R}'_\delta\right]\Gamma[-\mathcal{R}_\delta']\,\sin\left(-\frac{\pi}{2}\,\mathcal{R}'_\delta\right).
\end{equation}
Hence the integrated distribution is given by
\begin{align}
\sigma(\Delta)&=\sum_{\delta}\int\d\B_{\delta}\,\frac{\d\sigma_{0\delta}}{\d\B_{\delta}}\,\Xi_\B\,\frac{f_a(x_a,\Delta^2\,\mu_{\mathrm{f}}^2)\,f_b(x_b,\Delta^2\,\mu_{\mathrm{f}}^2)} {f_a(x_a,\mu_{\mathrm{f}}^2)\,f_b(x_b,\mu_{\mathrm{f}}^2)}\,\mathcal{S}_\delta(1/\Delta)\,\mathcal{C}_\delta(1/\Delta)\times \notag\\
&\times\exp\left[-\mathcal{R}_\delta(1/\Delta)-\gamma_E\,\mathcal{R}'_\delta(1/\Delta)\right]\frac{2}{\pi}\,\Gamma[-\mathcal{R}_\delta']\,\sin\left(-\frac{\pi}{2}\,\mathcal{R}'_\delta\right),\label{eq:final2}
\end{align}
with $f_i$ being the PDF of the incoming parton $i=\{a,b\}$, $x_i$ the momentum fraction carried by it, and $\mu_{\mathrm{f}}$ the factorisation scale.

The above expression for the integrated cross-section has a divergence when $\mathcal{R}_\delta'(1/\Delta)=1$. The divergence reaches values of $\Delta$ up to $0.3$ depending on the value of the jet $p_t$, and is always above $0.12$ for the quark channel contribution. This means that this approach is not well suited for phenomenological studies since the impact of the divergence on the distribution is severe.

In order to be able to avoid this divergence we follow another commonly used approach in which we perform the $b$ integral numerically. We proceed in the following way:
\begin{itemize}
\item We avoid the Landau-pole singularity in the radiator at $\bar{b}_{\max}=\exp[1/(2\,\alpha_s\,\beta_0)]$ by cutting off the $b$ integral at $\bar{b}=\bar{b}_{\mathrm{max}}$ setting the radiator to $\infty$ (and thus the integrand to zero) above this value.
\item We avoid low scales in the PDFs (which correspond to large values of $b$) by making the replacement $b\to b^*=b/\sqrt{1+b^2/b_{\mathrm{lim}}^2}$ in the radiator and PDFs \cite{Collins:1984kg}, where we choose $b_{\mathrm{lim}}=\mu_{\mathrm{f}}^2/Q_0^2$ and $Q_0\sim 1\,\mathrm{GeV}$ is a cutoff scale for PDFs.
\item At small $b$ the real and virtual contributions cancel each other and therefore the radiator should be equal to zero in the limit $b\to 0$. We thus freeze both the radiator and the scale of PDFs for values of $b$ that are less than 1.
\end{itemize}
Hence the expression for the cross-section to be numerically integrated is written as follows
\begin{align}
\sigma(\Delta)&=\sum_{\delta}\int\d\B_{\delta}\,\frac{\d\sigma_{0\delta}}{\d\B_{\delta}}\,\Xi_\B\left[\frac{2}{\pi}\,\mathrm{Si}(\bar{\Delta})
+\frac{2}{\pi}\int_{1}^{b_{\mathrm{max}}}\frac{\d b}{b}\sin(b\,\bar{\Delta})\times \right.\notag\\
&\left.\times \frac{f_a(x_a,\mu_{\mathrm{f}}^2/b^{*2})\,f_b(x_b,\mu_{\mathrm{f}}^2/b^{*2})} {f_a(x_a,\mu_{\mathrm{f}}^2)\,f_b(x_b,\mu_{\mathrm{f}}^2)}\,\mathcal{S}_\delta(b^*)\,\mathcal{C}_\delta(b^*)
\exp[-\mathcal{R}_{\delta}(b^*)] \right],\label{eq:final}
\end{align}
with $\mathrm{Si}$ standing for the Sine-integral function and $\bar{\Delta} = \Delta\,e^{-\gamma_E}$.

\subsection{Convolution with the Born cross-section}

To obtain the integrated distribution we generate a sample of parton-level Born events for the process of production of a $Z$ boson \footnote{The $Z$ boson is chosen not to decay. This only affects the total cross-section and does not impact the resummed distribution when it is normalised to the Born cross-section.} and a jet in proton-proton collisions, at centre-of-mass energy $\sqrt{s}=7\,\mathrm{TeV}$, using \texttt{MadGraph5} \cite{Maltoni:2002qb, Alwall:2014hca} in the ``\texttt{Les Houches Event File}'' format \cite{Alwall:2006yp}. We use CTEQ6L PDFs \cite{Pumplin:2002vw} interfaced to \texttt{MadGraph5} through \texttt{LHAPDF6} \cite{Buckley:2014ana} with fixed renormalisation and factorisation scales $\mu_{\mathrm{f}}=\mu_{\mathrm{r}}=150\,\mathrm{GeV}$. In the event generation we use the same experimental cuts $\Xi_\B$ as those employed by the CMS collaboration in ref. \cite{Chatrchyan:2013tna}. Specifically, the $Z$ boson is required to have $\tilde{p}_t>150\,\mathrm{GeV}$ and we require at least one jet with transverse momentum $p_t>50\,\mathrm{GeV}$ and rapidity $|y|<2.5$. The jets have radius $R=0.5$ and we use both $k_t$ and anti-$k_t$ clustering.

The generated events are then analysed using \texttt{MadAnalysis 5} \cite{Conte:2012fm}, where each Born event $\d\B_\delta$ is weighed by the integrand shown in the square brackets in eq. \eqref{eq:final} after performing the $b$ integration numerically using the \texttt{GSL-GNU Scientific Library}. The integrated distribution is then obtained by summing all the event weights and dividing by the effective luminosity $\mathcal{L}=N_{\mathrm{tot}}/\sigma_0$, with $N_{\mathrm{tot}}$ the total number of events and $\sigma_0$ the cross-section for the generated parton-level Born events. The differential distribution is then easily obtained by straightforward numerical differentiation.

\subsection{NNLL corrections at two loops}

At $\mathcal{O}(\alpha_s^n)$ in the perturbative expansion, the logarithm $\alpha_s^n L^{2n-2}$ is not fully accounted for by the resummation. We can incorporate such an NNLL term \footnote{In the \emph{perturbative expansion}, LL refers to $\alpha_s^nL^{2n}$, NLL refers to $\alpha_s^{n}L^{2n-1}$, and NNLL refers to $\alpha_s^nL^{2n-2}$. This hierarchy is different from the logarithmic accuracy in the \emph{exponent} of the resummed distribution.} by computing the NLO fixed-order correction term $\alpha_s\,C_1^{(\delta)}(\mathcal{B}_{\delta})$ for the channel $\delta$ and the differential Born configuration $\mathcal{B}_{\delta}$. To see this we note that the sub-leading logarithms $\alpha_s^n L^{n-1}$, not accounted for in the exponent of the resummed result, produce at most an $\mathcal{O}(\alpha_s^nL^{2n-3})$ term when expanded at fixed order, which means they contribute at NNNLL accuracy. On the other hand, the cross-talk of the constant term $\alpha_s\,C_1^{(\delta)}(\mathcal{B}_{\delta})$ with the expansion of the controlled double logarithms in the exponent $\alpha_s^n L^{2n}$ produce the NNLL terms $\alpha_s^{n+1} L^{2n}\sim \alpha_s^n L^{2n-2}$ \cite{Banfi:2010xy}.

This constant term may be computed by fully subtracting the large single ($\alpha_s\ln(1/\Delta)$) and double ($\alpha_s\ln^2(1/\Delta)$) logarithms from the full NLO Born-differential distribution \cite{Banfi:2010xy}
\begin{equation}
\alpha_s\,C_1^{(\delta)}(\mathcal{B}_{\delta})=\frac{1}{\d\sigma_{0\delta}/\d\mathcal{B}_{\delta}}\,\lim_{\Delta\to 0}\left[\int_{0}^{\Delta}
\frac{\d^2\sigma^{(\delta)}_{\mathrm{NLO}}}{\d\mathcal{B}_{\delta}\,\d\delta\phi}\,\d\delta\phi-\frac{\d\sigma_1^{(\delta)}(\Delta)}{\d\mathcal{B}_{\delta}}\right],
\end{equation}
where $\d^2\sigma^{(\delta)}_{\mathrm{NLO}}/\d\mathcal{B}_{\delta}\,\d\delta\phi$ is the NLO differential cross-section in both $\delta\phi$ and the Born configuration $\mathcal{B}_{\delta}$ for channel $\delta$, and $\d\sigma^{(\delta)}_1/\d\mathcal{B}_{\delta}$ is the contribution of the channel $\delta$ to the expansion of the differential (in $\mathcal{B}_{\delta}$) resummed distribution up to $\mathcal{O}(\alpha_s)$. The corrected integrated resummed distribution (in its formal form) is then written as
\begin{align}
\sigma(\Delta)&=\sum_{\delta}\int\d\B_{\delta}\,\frac{\d\sigma_{0\delta}}{\d\B_{\delta}}\,\Xi_\B\left(1+\alpha_s\,C_1^{(\delta)}(\mathcal{B}_{\delta})\right)\times\notag\\
&\times\frac{2}{\pi}\int_{0}^\infty\frac{\d b}{b}\sin(b\,\Delta)\,\frac{f_a(x_a,\mu_{\mathrm{f}}^2/\bar{b}^2)\,f_b(x_b,\mu_{\mathrm{f}}^2/\bar{b}^2)} {f_a(x_a,\mu_{\mathrm{f}}^2)\,f_b(x_b,\mu_{\mathrm{f}}^2)}\,\mathcal{S}_{\delta}(\bar{b})\,\mathcal{C}_{\delta}(\bar{b})\exp\left[-\mathcal{R}_\delta(\bar{b})\right].
\label{eq:master3}
\end{align}

The computation of the Born-differential NLO cross-section (either numerically or analytically) is complicated. For the purpose of this paper, we define $\alpha_s\,C_1^{(\delta)}$ to be the average value over the Born configuration $\mathcal{B}_{\delta}$
\begin{equation}
\left\langle\alpha_s\,C_1^{(\delta)}\right\rangle = \frac{1}{\sigma_{0\delta}} \int\d\B_{\delta}\,\frac{\d\sigma_{0\delta}}{\d\mathcal{B}_{\delta}}\,\Xi_{\mathcal{B}}\,\alpha_s\,C_1^{(\delta)}(\mathcal{B})\,,
\end{equation}
and replace $\alpha_s\,C_1^{(\delta)}(\mathcal{B})$ by its average value $\left\langle\alpha_s\,C_1^{(\delta)}\right\rangle\equiv \alpha_s\,C_1^{(\delta)}$ for each channel.

\subsection{Comparison to MC results at fixed order}

One can verify that the resummed distribution correctly captures the large single and double logarithms when expanded at $\mathcal{O}(\alpha_s)$ by comparing it with the results obtained from fixed-order programs such as \texttt{MadGraph5\_aMC@NLO} and \texttt{MCFM}. The expanded distribution at $\mathcal{O}(\alpha_s)$ is given by
\begin{align}
\sigma_1(\Delta)&=\sum_{\delta}\int\d\B_{\delta}\,\frac{\d\sigma_{0\delta}}{\d\B_{\delta}}\,\Xi_\B
\left(1+G^{(\delta)}_{11}\,\frac{\alpha_s}{\pi}\ln\frac{1}{\Delta}+G^{(\delta)}_{12}\,\frac{\alpha_s}{\pi}\ln^2\frac{1}{\Delta}\right).\label{eq:expansion0}
\end{align}
The expressions of the coefficients $G^{(\delta)}_{11}$ and $G^{(\delta)}_{12}$ are given in appendix \ref{sec:AppB}. We perform the integration over the Born configuration $\d\mathcal{B}_{\delta}$ exactly as described in the previous subsections.

\begin{figure}[ht]
\centering
\includegraphics[width=0.55\textwidth]{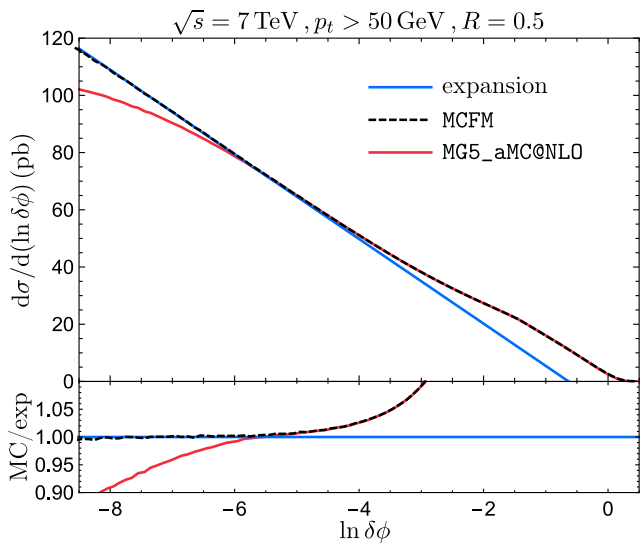}\hfil
\caption{\label{fig:05} Comparison between the differential distributions $\d\sigma/\d(\ln\delta\phi)$ obtained by expanding the resummed distribution up to $\mathcal{O}(\alpha_s)$ and fixed-order programs \texttt{MadGraph5\_aMC@NLO} and \texttt{MCFM}.}
\end{figure}
We show in figure \ref{fig:05} a comparison between the differential expanded distribution (obtained by differentiating eq. \eqref{eq:expansion0} with respect to $\ln\Delta$) and the fixed-order MC distribution obtained with \texttt{MadGraph5\_aMC@NLO} and \texttt{MCFM}. At NLO the channel separation becomes non-trivial due to the presence of other sub-processes, such as $qq\to qq$. 

Three things to observe in the plots. First, the \texttt{MCFM} and expansion results agree up to values of $\ln\delta\phi$ around $-4.5$ ($\delta\phi \sim 0.01$). At larger values of $\delta\phi$ the logarithms are not enhanced and NLO effects, not accounted for by resummation, dominate the MC distribution. Second, the \texttt{MadGraph5\_aMC@NLO} distribution agrees with \texttt{MCFM} result over the entire range of $\delta\phi$ down to $\ln\delta\phi$ around $-6$ ($\delta\phi\sim0.003$), where deviations due to integration precision are expected to affect the distribution. Third, there is a kinematical cutoff in the exact NLO MC distribution at around $\delta\phi=1.54$ (slightly less than $\pi/2$). This cutoff depends on the value of the cut on $\tilde{p}_t$ of the $Z$ boson, and may be shown to be given by the expression
\begin{equation}
\cos\delta\phi^{\mathrm{max}}=\frac{\tilde{p}_{t,\mathrm{cut}}}{\sqrt{s}-\sqrt{\tilde{p}^{2}_{t,\mathrm{cut}}+m_Z^2}}\approx \frac{\tilde{p}_{t,\mathrm{cut}}}{\sqrt{s}}\,,
\end{equation}
where $m_Z$ is the $Z$ boson mass. For $\sqrt{s}=7\,\mathrm{TeV}$ and $\tilde{p}_{t,\mathrm{cut}}=150\,\mathrm{GeV}$ this yields $\delta\phi^{\mathrm{max}} = 1.549$.

\section{Comparison to parton showers and experimental data}

In this section we present our results for the resummed distribution $1/\sigma\,\d\sigma/\d\delta\phi$ and compare our findings with parton shower results obtained with various MC event generators and with experimental data from the CMS collaboration \cite{Chatrchyan:2013tna}. In the remainder of this section we restrict ourselves to showing preliminary results and comparisons, while we leave other phenomenological studies, namely matching to fixed order, non-perturbative effects, and uncertainties of the distribution, to our forthcoming work.

\begin{figure}[ht]
\centering
\includegraphics[width=0.51\textwidth]{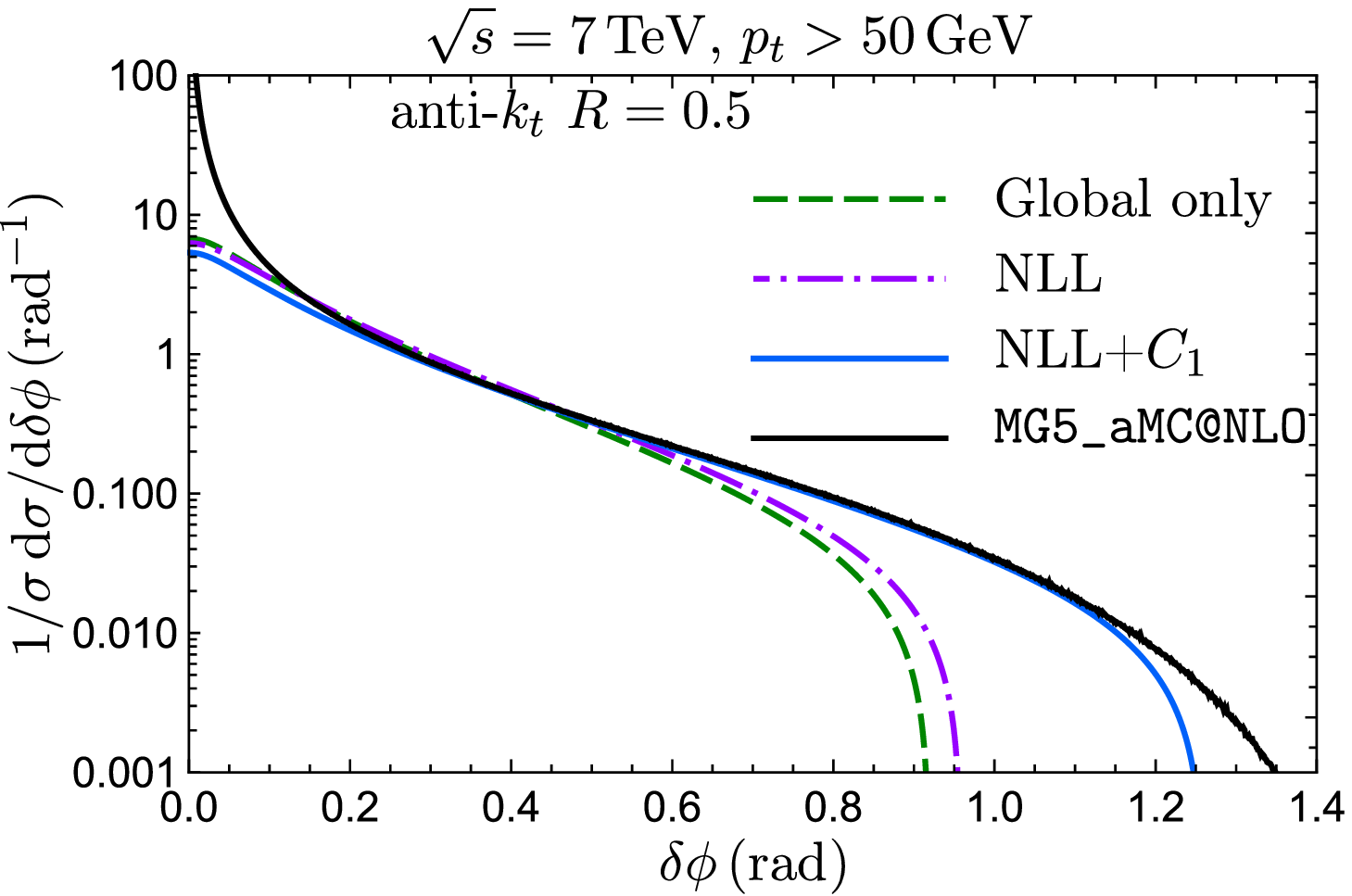}\hfil
\includegraphics[width=0.47\textwidth]{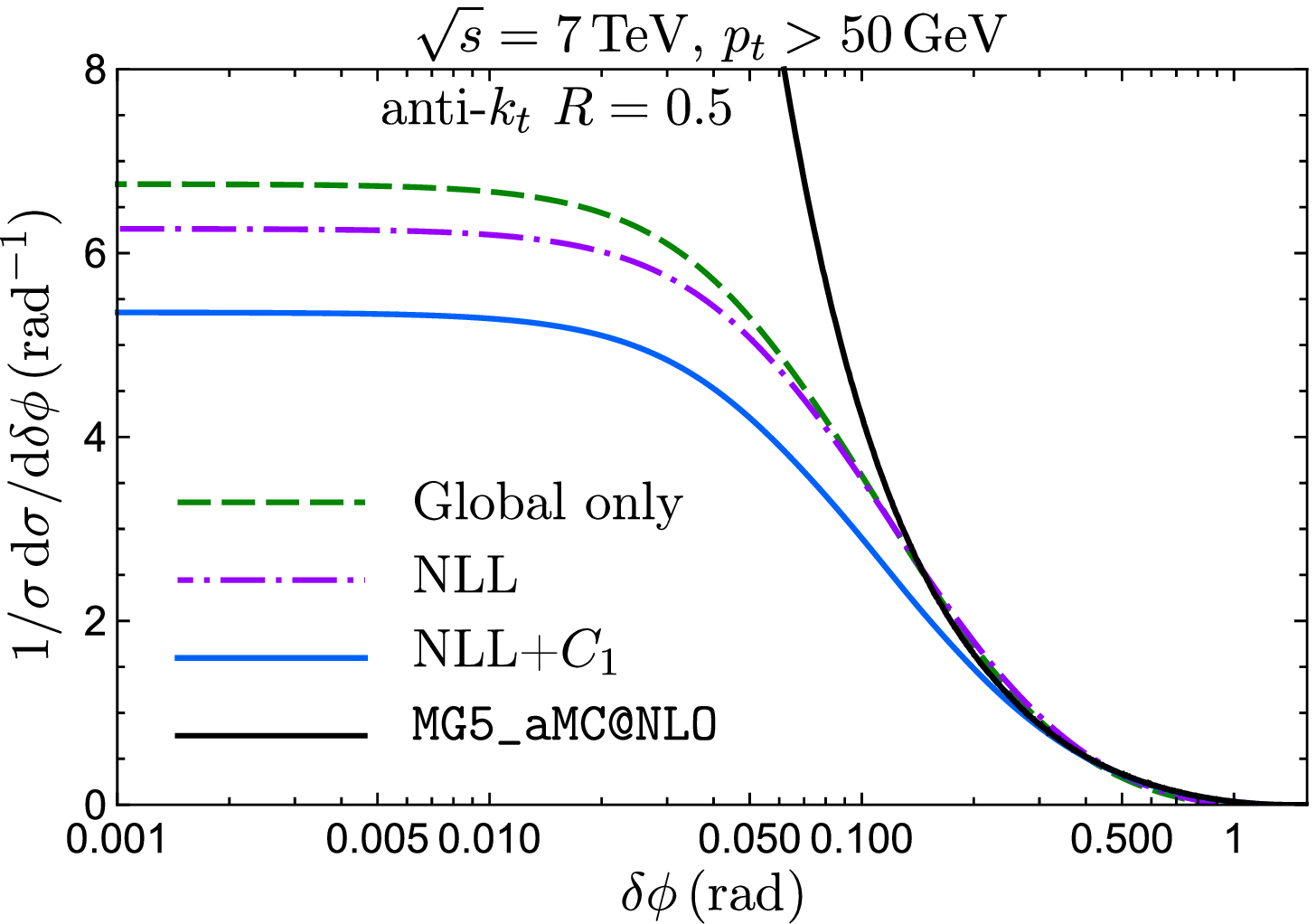}\\
\includegraphics[width=0.51\textwidth]{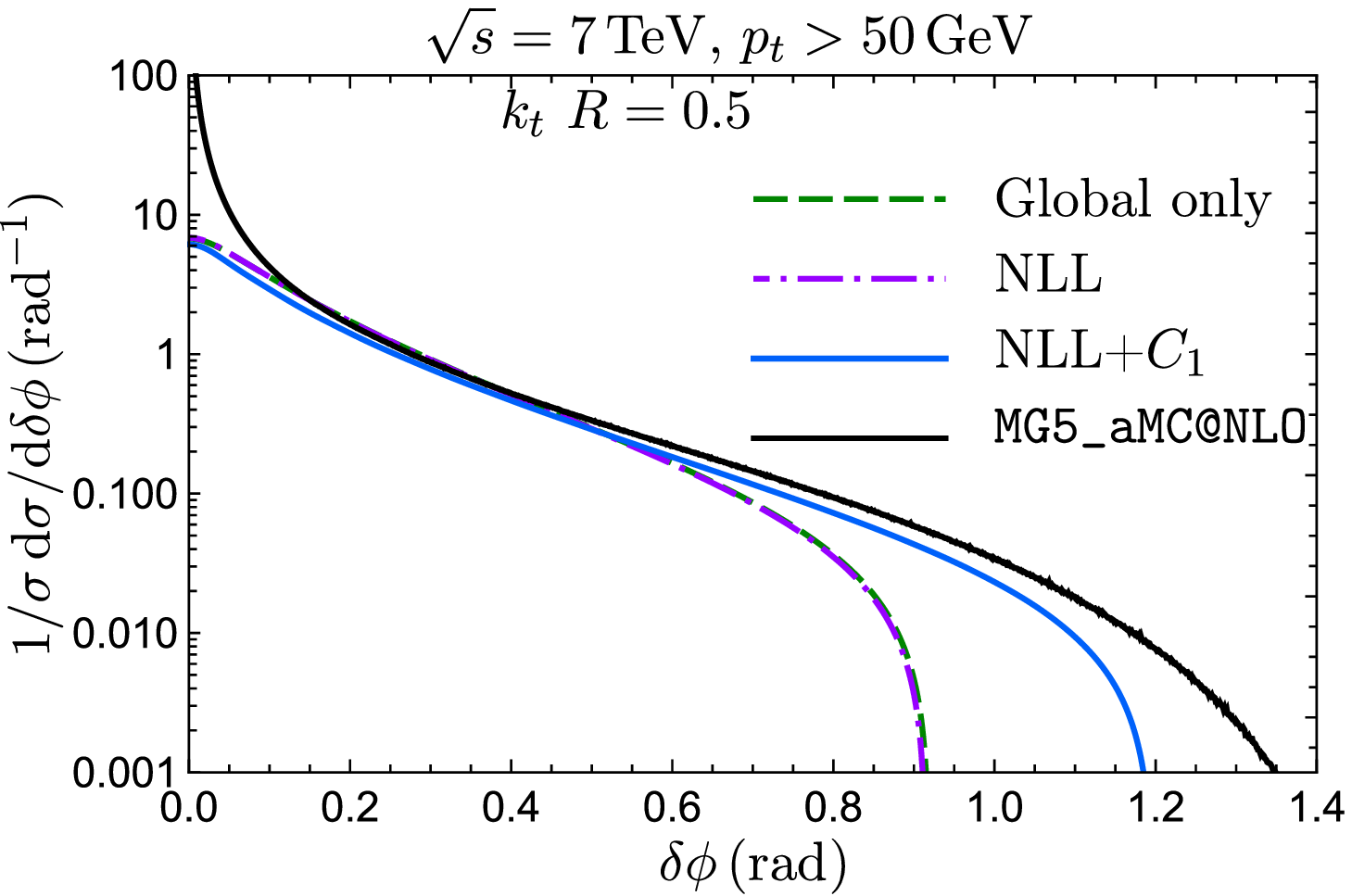}\hfil
\includegraphics[width=0.47\textwidth]{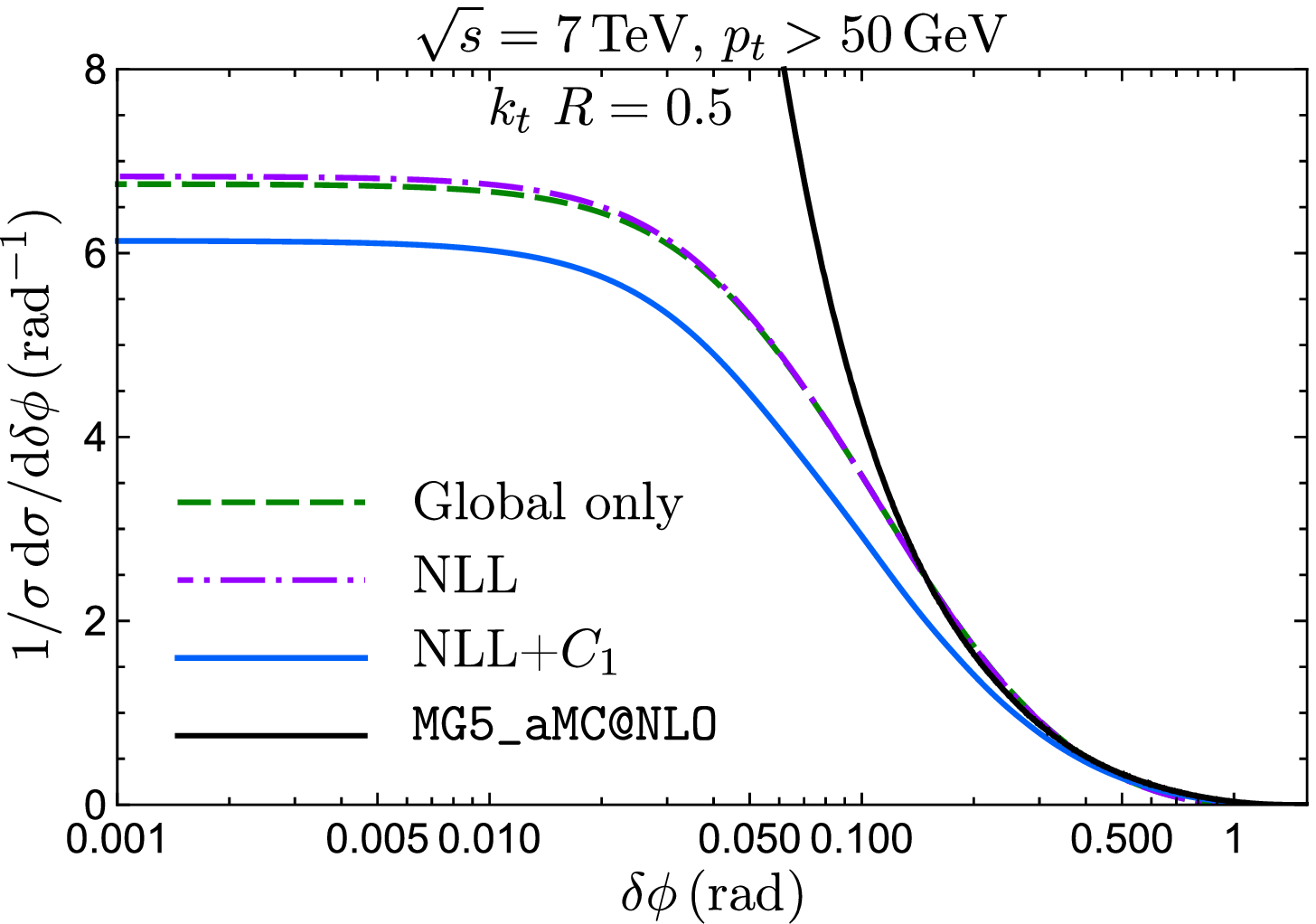}
\caption{\label{fig:06} The differential distribution $1/\sigma\,\d\sigma/\d\delta\phi$ with anti-$k_t$ clustering (top) and $k_t$ clustering (bottom). We show here the global resummed distribution without NGLs/CLs, the full NLL resummed distribution (including NGLs/CLs), and the corrected NLL distribution including the constant $C_1$. Also shown is the fixed-order MC result at NLO.}
\end{figure}

We show in figure \ref{fig:06} plots of the resummed global distribution (without NGLs/CLs), the resummed distribution in the $k_t$ and anti-$k_t$ algorithms (with CLs/NGLs), and the corrected resummed distribution for NNLL effects at fixed order (including the $C_1$ constant). We also show, in the same figure, a plot of the fixed-order MC distribution obtained with \texttt{MadGraph5\_aMC@NLO}.
For small values of $\delta\phi$ the resummed distribution tends towards a constant. This observation may be interpreted, as mentioned earlier, by the fact that the very small values of $\delta\phi$ may be generated by vectorial cancellation of hard emissions which takes over from Sudakov suppression of soft emissions.

In the anti-$k_t$ algorithm and for the value of the jet radius $R=0.5$ we see that the impact of NGLs on the distribution is important at small and moderate values of $\delta\phi$. Additionally, including the constant $C_1$ impacts the distribution both in the small-$\delta\phi$ region as well as in the tail of the distribution, bringing it to a better matching with the fixed-order MC result for values of $\delta\phi$ up to order $1$. As for $k_t$ clustering, we note that the impact of NGLs and CLs is not significant for all values of $\delta\phi$, and that including the $C_1$ constant improves the distribution both at small and large $\delta\phi$, but one still needs the matching in order to obtain a better behaviour at the tail of the distribution. The fixed-order MC distribution behaves reasonably well down to values of $\delta\phi$ of order $0.1$, where it starts to behave in a logarithmically divergent way.

As mentioned earlier, at NLO the MC distribution has a kinematical cutoff on the observable around $\pi/2$, when there is only one extra hard emission not clustered to the ``measured'' jet. However, beyond NLO the distribution receives contributions from two or more extra hard emissions giving possible values of $\delta\phi$ up to $\pi$. It therefore makes no sense to perform a matching of the resummed distribution to NLO without including effects of higher jet multiplicities. The matching procedure itself is complicated by the fact that the resummation is performed in $b$ space and we have no analytical form of the resummed distribution in $\delta\phi$ space.

\begin{figure}[ht]
\centering
\includegraphics[width=0.51\textwidth]{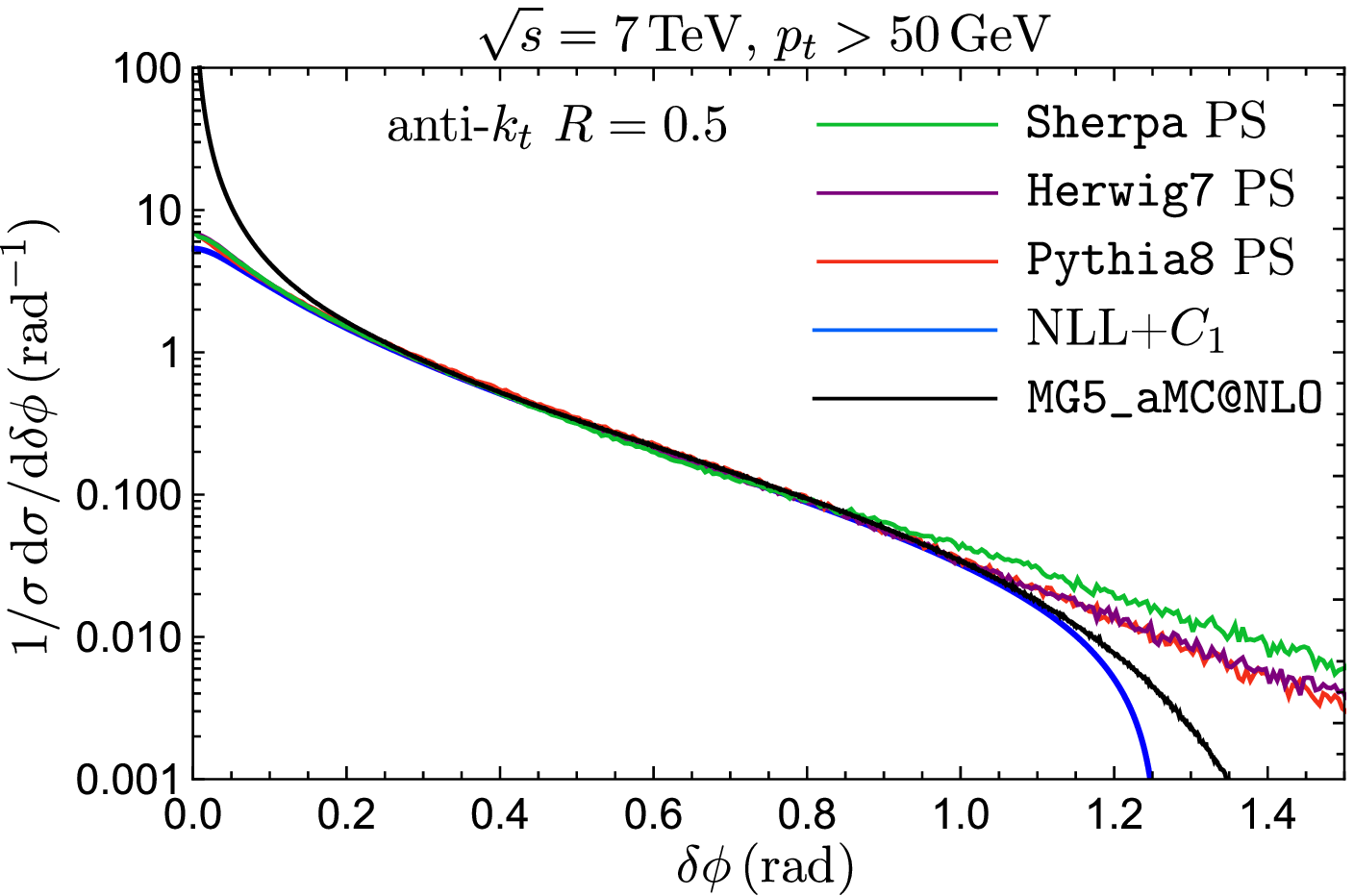}\hfil
\includegraphics[width=0.47\textwidth]{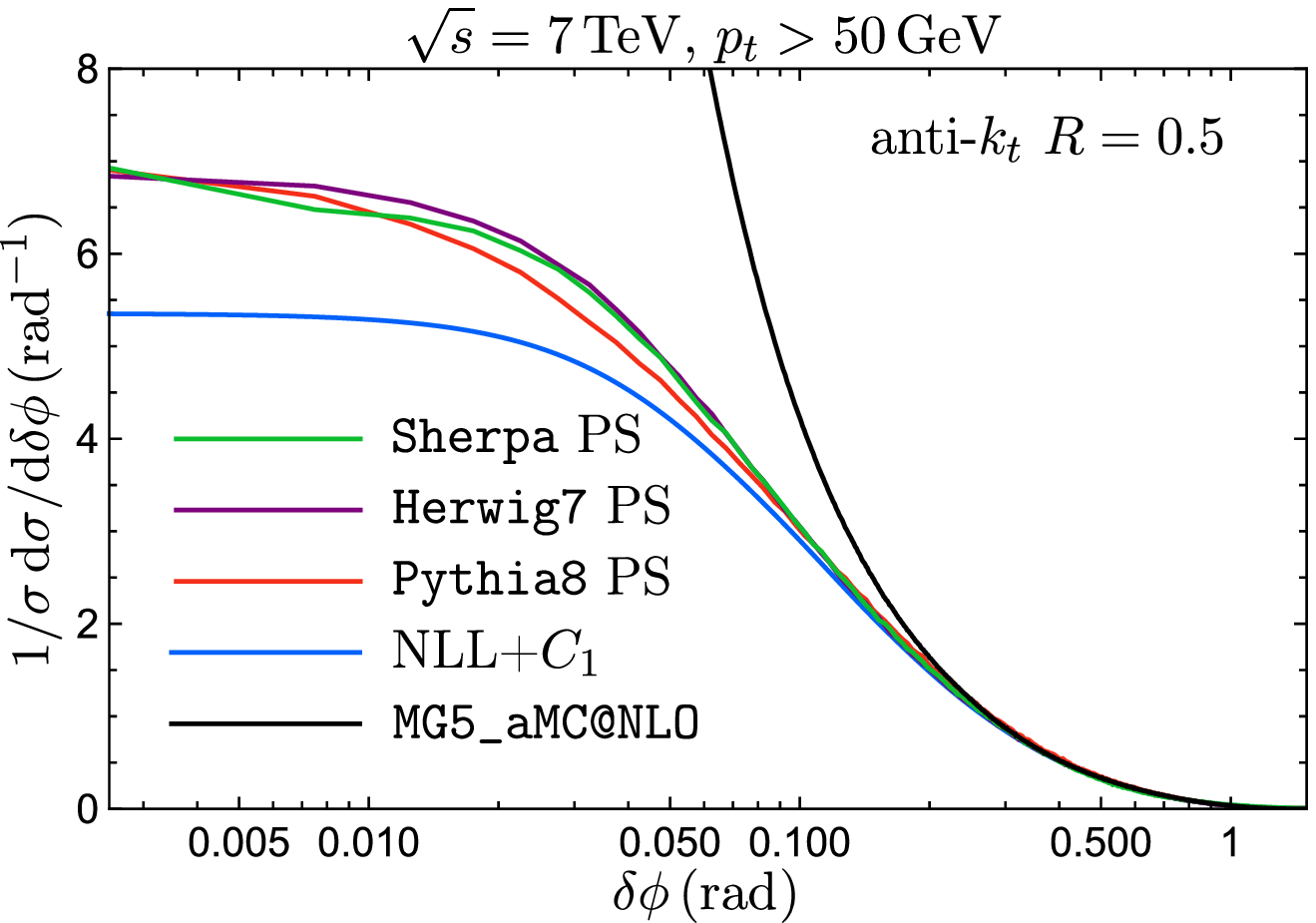}\\
\includegraphics[width=0.51\textwidth]{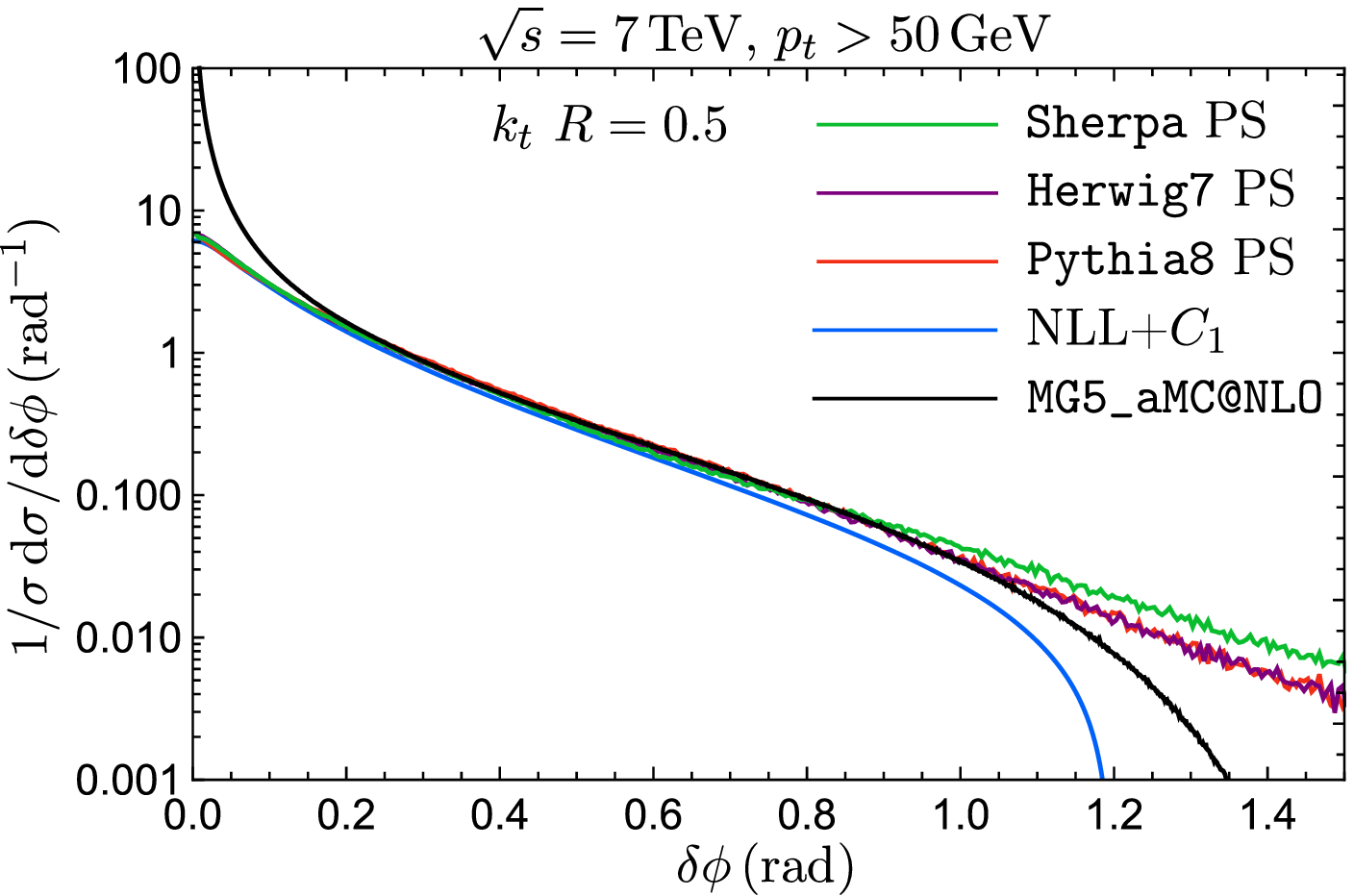}\hfil
\includegraphics[width=0.47\textwidth]{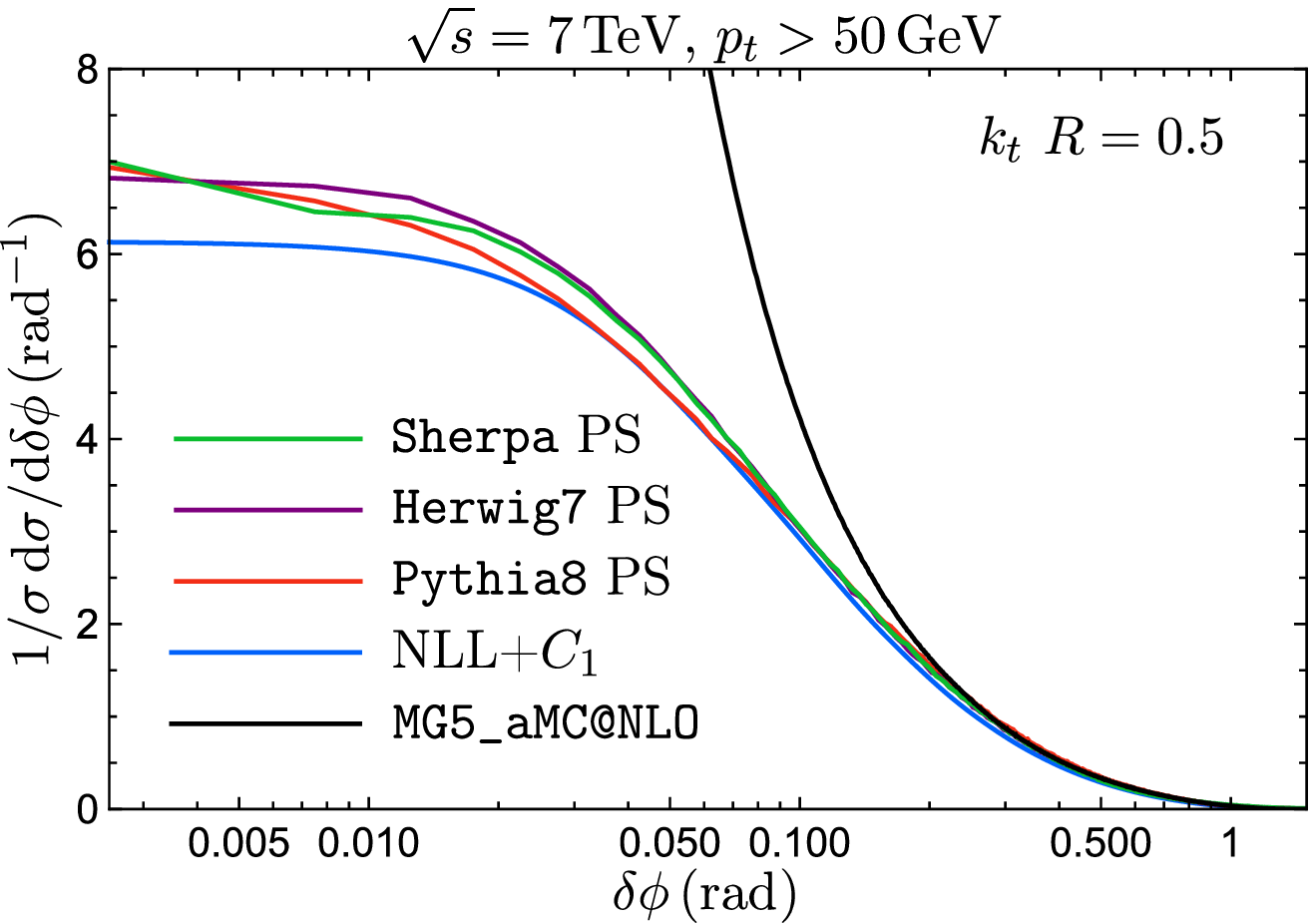}
\caption{\label{fig:07} The differential distributions $1/\sigma\,\d\sigma/\d\delta\phi$ with anti-$k_t$ clustering (top) and $k_t$ clustering (bottom). Results are shown for the resummed distribution NLL+$C_1$ and parton showers (PS) obtained with various MC event generators. Also shown is the fixed-order NLO MC result.}
\end{figure}

We show in figure \ref{fig:07} a comparison of our resummed distribution with parton shower results obtained with \texttt{Pythia 8}  \cite{Sjostrand:2014zea,Alwall:2008qv}, \texttt{Herwig++} \cite{Bahr:2008pv,Bellm:2015jjp} and \texttt{Sherpa} \cite{Sherpa:2019gpd}. The \texttt{Pythia 8} and \texttt{Herwig++} results are obtained by showering \texttt{MadGraph5\_aMC@NLO} events, while the \texttt{Sherpa} results are obtained using the stand-alone version. In all cases the jets are clustered with \texttt{FastJet} \cite{Cacciari:2011ma}. We also include in the plots the NLO results from \texttt{MadGraph5\_aMC@NLO}.

We note that the parton-shower results are in agreement with the resummed result for intermediate values of $\delta\phi$ and with fixed-order NLO results from \texttt{MadGraph5\_aMC@NLO}. In the parton shower results the observable continues to have a distribution for values of $\delta\phi$ up to $\pi$ as explained above, and deviation from our NLL-resummed result starts at around $1.0$ at the tail of the distribution, where also fixed-order NLO results start to make no sense.

At small values of $\delta\phi$ our resummed distribution in the anti-$k_t$ algorithm has a lower value than the parton shower results. It would seem that the global result is performing better than the NLL+$C_1$ result in this region when compared with these parton shower results. On the contrary, one can make the observation that \texttt{Pythia 8} results in the small-$\delta\phi$ region are quite high compared to other results obtained in previous works. For instance, in ref. \cite{Chien:2020hzh} it was noted that the parton shower result, obtained with \texttt{Pythia 8} stand-alone and multiplied by an NLO $K$ factor of 1.6 in the WTA recombination scheme, is also slightly larger than the resummed result at NLL+NLO and NNLL+NLO accuracy for small azimuthal decorrelation. A similar observation may be noted in the work of the CMS collaboration in ref. \cite{Chatrchyan:2013tna}, where the \texttt{Pythia 8} stand-alone result (which includes non-perturbative effects) has a higher value than the experimental data in the last bin corresponding to small azimuthal decorrelation.  We have verified that this is not due to non-perturbative hadronisation and underlying-event effects, which have a small impact on the distribution as was too observed in ref. \cite{Chien:2020hzh}.

\begin{figure}[ht]
\centering
\includegraphics[width=0.57\textwidth]{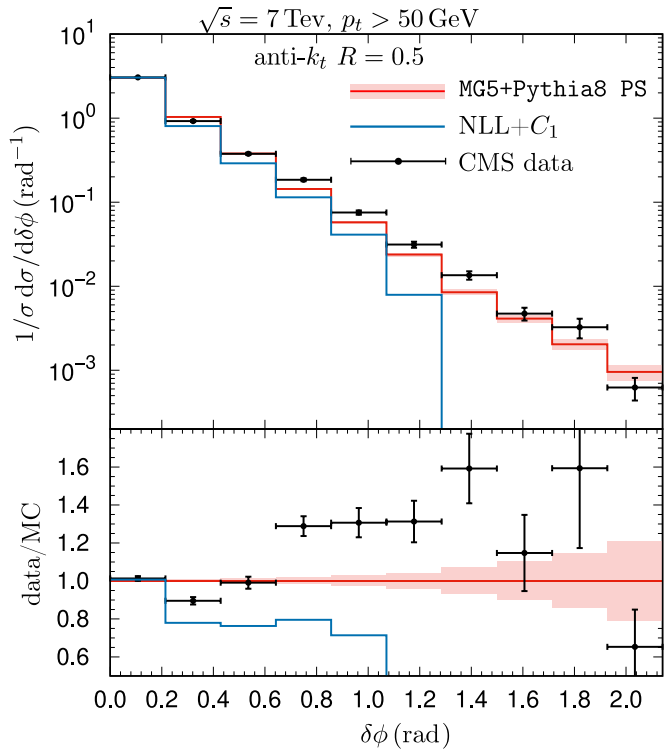}
\caption{\label{fig:08} Comparison between the NLL-resummed distribution, the \texttt{Pythia 8} parton shower result (including $Z$ + 2 jet process), and the CMS data.}
\end{figure}

The discrepancy is explained by the fact that the parton shower results do not include effects of higher jet multiplicities. If one includes such effects, the tail of the distribution becomes higher due to contributions from 2 and 3 jet events, and at the same time the small-$\delta\phi$ region gets lower due to the normalisation of the distribution. \footnote{The parton shower and experimental distributions are normalised such that the area under the curve is equal to 1. As the distribution raises in the tail it gets lower at small $\delta\phi$ to preserve the area under the distribution.} One can include such effects in \texttt{MadGraph5\_aMC@NLO} by merging the process of production of $Z$ + 2 jets (and also possibly $Z$ + 3 jets) with the Born process $Z$ + jet, and performing the showering of events with the said MC parton showers. Doing so gives a reasonable agreement with our resummed distribution and with experimental data in the last bin (corresponding to small $\delta\phi$), as observed from figure \ref{fig:08}.

\section{Conclusions}

The azimuthal decorrelation is an important jet observable that plays an indispensable role in paving the way for a better understanding of QCD as well as accurate background-signal discrimination. We have presented in the current work a full NLL resummation of this quantity that includes the subtle NGLs and CLs contributions.

After a detailed computation of the global resummed form factor with full jet-radius dependence, we clarified the structure of NGLs and CLs at two-loops order with full colour and jet-radius dependence, and for both anti-$k_t$ and $k_t$ jet algorithms. The reduction of NGLs with $k_t$ clustering is, as has previously been reported, seen. Jets initiated by gluons generally have higher NGLs and CLs coefficients compared to those initiated by quarks. This is mainly due to the large colour factor of the former. The all-orders treatment of the said logarithms in the $k_t$ algorithm has been confined to the exponential of the two-loops result due to a) the reasonable approximation of the latter exponential, and b) the difficulties associated with obtaining the resummed NGLs/CLs to all orders. For the anti-$k_t$ algorithm we numerically computed the resummed NGLs to all orders in the large-$\Nc$ limit.

Comparisons to both fixed-order and parton-shower MC programs have been performed. Various factors have been shown to play important roles in order to achieve reasonable agreement between our semi-analytical findings and those of the MC programs. These noticeably include the value of the jet radius, NGLs and/or CLs contributions, and the constant term $C_1$. Discrepancies with some of the MC parton showers at low values of the azimuthal decorrelation observable have been ascribed to the necessity of inclusion of higher jet multiplicities. This has been confirmed particularly via comparisons to experimental data from the CMS collaboration.

We have restricted ourselves, in the present work, to pure perturbative treatment. Non-perturbative corrections, fixed-order matching and statistical and systematic uncertainties will be considered in a future work. We finally note that our calculations may easily be extended to the process of production of a $W$, $\gamma$, or Higgs boson in association with a jet. The QCD calculations are exactly identical in these cases, with the only differences being the Born cross-section and the opening of the channel $gg\to g$ in the case of Higgs + jet production.

\appendix

\section{Radiator}

\label{sec:rad}

\subsection{In-in dipole}

We begin by evaluating the integral for the $(ab)$-dipole contribution to the radiator, given by
\begin{equation}
\mathcal{R}_{\delta}^{(ab)}(\bar{b})=\mathcal{C}_{ab}\int\frac{\as(\kappa_{t,ab}^2)}{\pi}\,\frac{\d k_t}{k_t}\,\d\eta\,\frac{\d\phi}{2\pi}\,\Theta_{\out}(k)\,w_{ab}^k\,\Theta\left(k_{t}|\sin\phi|-p_t/\bar{b}\right).
\end{equation}
For this dipole we simply have $w_{ab}^k=1$ and $\kappa_{t,ab}^2=k_t^2$. To achieve the evaluation of this integral we write the step function $\Theta_{\out}(k)=1-\Theta_{\inn}(k)$, and perform the integration over each term separately.

\subsubsection{Integrating the whole phase space}
\label{sec:wps}
We evaluate the integral
\begin{equation}
\mathcal{R}_{\delta}^{(ab)-\mathrm{all}}(\bar{b})=\mathcal{C}_{ab}\int\frac{\as(k_t^2)}{\pi}\,\frac{\d k_t}{k_t}\,\d\eta\,\frac{\d\phi}{2\pi}\,\Theta\left(k_{t}|\sin\phi|-p_t/\bar{b}\right).
\end{equation}
The rapidity integration has a cutoff originating from the requirement that the emitted gluon when collinear to one of the incoming legs must have an energy less than that of the emitting parent hard parton, i.e. $k_t\cosh\eta< x_a\sqrt{s}/2$ and $k_{t}\cosh \eta< x_b\sqrt{s}/2$. When the emission is collinear to the leg $(a)$ we have $k_t\cosh \eta\approx k_te^\eta/2 < x_a \sqrt{s}/2$, and similarly when the emission is collinear to the leg $(b)$ we obtain $k_te^{-\eta}/2 < x_b \sqrt{s}/2$. Including contributions from hard-collinear emissions to the corresponding hard leg is simply achieved by making the replacement $x_a\sqrt{s}\to x_a \sqrt{s}\,e^{B_a}$ and $x_b\sqrt{s}\to x_b\sqrt{s}\,e^{B_b}$, where
\begin{equation}
\begin{aligned}
B_q &= -\frac{3}{4}                                           &&\text{for quark legs}\,, \\
B_g &= -\frac{11\,\CA-4\,\mathrm{T_R}\,\mathrm{n_f}}{12\,\CA}=-\frac{\pi\beta_0}{\CA} \quad&&\text{for gluon legs}\,,
\end{aligned}
\end{equation}
where $\mathrm{T_R}=1/2$ is the normalisation constant for the SU$(\Nc)$ generators, $\mathrm{n_f}=5$ is the number of active quark flavours, and $\beta_0$ is the one-loop coefficient of the QCD beta function (defined below). Additionally, the scale of the PDFs is changed from $\mu_{\mathrm{f}}^2 \to \mu_{\mathrm{f}}^2/\bar{b}^2$. We thus deduce that
\begin{equation}
-\ln\left(\frac{x_b}{k_t}{\sqrt{s}}\,e^{B_b}\right)<\eta<\ln \left(\frac{x_a}{k_t}\sqrt{s}\,e^{B_a}\right).
\end{equation}
Performing the integration over $\eta$ we obtain
\begin{equation}
\int \d\eta = \ln \frac{x_a\,x_b\,s}{k_t^2}+B_a+B_b=\ln\frac{Q_{ab}^2}{k_t^2}+B_a+B_b\,,
\end{equation}
where we define $Q^2_{ij}=(p_i+p_j)^2=2\,p_i\cdot p_j$. The latter quantities are related to the partonic Mandelstam variables, $Q_{ab}^2=\hat{s}=x_a\,x_b\,s$, $Q_{aj}^2=-\hat{t}=x_a\sqrt{s}\,p_t\,e^{-y}$, and $Q_{bj}^2=-\hat{u} = x_b\,\sqrt{s}p_t\,e^y$\,.

Next, we perform the integration over $k_t$. Using the two-loops QCD beta function we have
\begin{equation}
\as(k_{t}^2)=\as\left[\frac{1}{1+2\,\as\,\beta_0\ln(k_t/p_t)}-\frac{\beta_1}{\beta_0}\,\as\,\frac{\ln\left[1+2\,\as\,\beta_0\ln(k_t/p_t)\right]}{[1+2\,\as\,\beta_0\ln(k_t/p_t)]^2}\right],
\end{equation}
where $\as$ in the right-hand-side is evaluated at the renormalisation scale $\mu_\mathrm{r}=p_t$, and $\beta_0$ and $\beta_1$ are the one and two-loops coefficients of the QCD beta function given by
\begin{equation}
\begin{split}
\beta_0&=\frac{11\,\mathrm{C_A}-2\,\mathrm{n_f}}{12\,\pi}\,,\\
\beta_1&=\frac{17\,\mathrm{C_A^2}-5\,\mathrm{C_A}\,\mathrm{n_f}-3\,\mathrm{C_F}\,\mathrm{n_f}}{24\pi^2}\,.
\end{split}
\end{equation}
Integrating over $k_t$ keeping only up to single-logarithmic terms we arrive at
\begin{align}
&\int_{p_t/(\bar{b}\,|\sin\phi|)}^{p_t}\frac{\as(k_t^2)}{\pi}\,\frac{\d k_t}{k_t}\left(B_a+B_b+\ln\frac{Q_{ab}^2}{p_t^2}-\ln\frac{k_t^2}{p_t^2} \right)=\notag\\
&=-L\,\frac{1}{2\pi\beta_0\,\Lambda}\left[2\Lambda+\ln(1-2\Lambda)\right]-\frac{\beta_1}{2\pi\beta_0^3}\left[\frac{1}{2}\ln^2(1-2\Lambda)+\frac{\ln(1-2\Lambda)+2\Lambda}{1-2\Lambda}\right]+\notag\\
&+\frac{1}{\pi\beta_0}\ln|\sin\phi|\,\frac{2\Lambda}{1-2\Lambda}-\frac{1}{2\pi\beta_0}\left[\ln\frac{Q_{ab}^2}{p_t^2}+B_a+B_b\right]\ln(1-2\Lambda)\,,
\end{align}
with $L=\ln\bar{b}$ and $\Lambda=\as\beta_0\,L$.

Finally, we perform the azimuthal averaging using $\frac{1}{2\pi}\int_0^{2\pi} \ln|\sin\phi|\,\d\phi=-\ln2$, which simply results in changing $\ln|\sin\phi| \to -\ln 2$\,.

\subsubsection{Subtracting the jet region}

Here we introduce the change of variables
\begin{subequations}\label{eq:change}
\begin{align}
\eta-y&=R\,r\cos\theta\,,\\
\phi-\pi&=R\,r\sin\theta\,.
\end{align}
\end{subequations}
The step function restricting the gluon to be in the jet is $\Theta[R^2-(\eta-y)^2-(\phi-\pi)^2]=\Theta(1-r^2)$. The angular phase space becomes $\d\eta\,\d\phi=R^2\,r\,\d r\,\d\theta$. We thus evaluate the integral
\begin{equation}
\mathcal{R}_{\delta}^{(ab)-\mathrm{in}}(\bar{b})=\mathcal{C}_{ab}\,R^2\int\frac{\as(k_t^2)}{\pi}\,\frac{\d k_t}{k_t}\,r\,\d r\,\frac{\d\theta}{2\pi}\,\Theta(1-r^2)\,\Theta\left(k_t|\sin(R\,r\,\sin\theta)|-p_t/\bar{b}\right).
\end{equation}
This term is free from collinear logarithms since it integrates only over emissions inside the outgoing jet away from the emitting incoming dipole. We thus use just the one-loop running of the coupling and expand at single logarithmic accuracy $\Theta\left(k_{t}|\sin(R\,r\,\sin\theta)|-p_t/\bar{b}\right)=\Theta\left(k_t-p_t/\bar{b}\right)$. It is then straightforward to obtain the result
\begin{equation}
\int_{p_t/\bar{b}}^{p_t}\frac{\as(k_t^2)}{\pi}\,\frac{\d k_t}{k_t}=\int_{p_t/\bar{b}}^{p_t}\,\frac{\as}{\pi}\,\frac{\d k_t}{k_t}\frac{1}{1+2\,\as\,\beta_0\,\ln(k_t/p_t)}
=-\frac{1}{2\pi\beta_0}\ln[1-2\Lambda]\,,
\end{equation}
where we discard sub-leading terms. The remaining angular integrations are straightforward and we obtain
\begin{equation}\label{eq:jetfunction}
\mathcal{R}_{\delta}^{(ab)-\mathrm{in}}(\bar{b})=-\mathcal{C}_{ab}\,\frac{1}{2}\,R^2\,\frac{1}{2\pi\beta_0}\ln[1-2\Lambda]\,.
\end{equation}

The overall contribution of the in-in dipole to the radiator is then given by
\begin{align}
\mathcal{R}_{\delta}^{(ab)}(\bar{b})&=-\mathcal{C}_{ab}\,L\,\frac{1}{2\pi\beta_0\,\Lambda}\left[2\Lambda+\ln(1-2\Lambda)\right]+\mathcal{C}_{ab}\,\frac{K}{4\pi^2\beta_0^2}\left[\ln(1-2\Lambda)+\frac{2\Lambda}{1-2\Lambda}\right]-\notag\\
&-\mathcal{C}_{ab}\,\frac{\beta_1}{2\pi\beta_0^3}\left[\frac{1}{2}\ln^2(1-2\Lambda)+\frac{\ln(1-2\Lambda)+2\Lambda}{1-2\Lambda}\right]-\mathcal{C}_{ab}\,\ln2\,\frac{1}{\pi\beta_0}\,\frac{2\Lambda}{1-2\Lambda}-\notag\\
&-\mathcal{C}_{ab}\frac{1}{2\pi\beta_0}\left[\ln \frac{Q_{ab}^2}{p_t^2}+B_a+B_b\right]\ln(1-2\Lambda)+\mathcal{C}_{ab}\,\frac{1}{2}\,R^2\,\frac{1}{2\pi\beta_0}\ln(1-2\Lambda)\,,
\end{align}
where we changed the coupling from the CMW \cite{Catani:1990rr} to the $\overline{\mathrm{MS}}$ renormalisation scheme by making the replacement
\begin{equation}
\alpha_{s,\mathrm{CMW}}=\alpha_{s,\overline{\mathrm{MS}}}+\alpha_{s,\overline{\mathrm{MS}}}^2\,\frac{K}{2\pi}\,,
\end{equation}
with
\begin{equation}
K=\CA\left(\frac{67}{18}-\frac{\pi^2}{6}\right)-\frac{5}{9}\,\mathrm{n_f}\,.
\end{equation}

\subsection{In-jet dipole}

Next we evaluate the integral for the contribution of the dipole $(aj)$ to the radiator
\begin{equation}
\mathcal{R}_{\delta}^{(aj)}(\bar{b})=\mathcal{C}_{aj}\int\frac{\as(\kappa_t^2)}{\pi}\,\frac{\d k_t}{k_t}\,\d\eta\,\frac{\d\phi}{2\pi}\,\Theta_{\out}(k)\,w_{aj}^k\,\Theta\left(k_t|\sin\phi|-p_t/\bar{b}\right),
\end{equation}
where we have for this dipole $\kappa_t=k_t/\sqrt{w_{aj}^{k}}$, with
\begin{equation}
w_{aj}^{k}=\frac{1}{2}\,\frac{\exp(\eta-y)}{\cosh(\eta-y)-\cos(\phi-\pi)}\,.
\end{equation}
We make the change of variable $k_t \to \kappa_t$, where now the integration over $\kappa_t$ is restricted in the range
\begin{equation}
\frac{p_t}{\bar{b}\,|\sin\phi|\sqrt{w_{aj}^{k}}}<\kappa_t<\frac{p_t}{\sqrt{w_{aj}^{k}}}\,.
\end{equation}

We expose the singularities resulting from collinear emissions to the dipole legs by writing
\begin{equation}
w_{aj}^{k}\,\Theta_{\out}(k)=\left[\omega_{aj}^{k}-\frac{1}{R^2\,r^2}\right]+\left[\frac{1}{R^2\,r^2}-\omega_{aj}^{k}\right]\Theta_{\inn}(k)+\frac{1}{R^2\,r^2}\,\Theta_{\out}(k)\,,
\end{equation}
where the variable $r$ has been introduced in eq. \eqref{eq:change}. In the first term the collinear pole to the outgoing jet ($r\to 0$) is subtracted from the antenna function and the integration is performed in the entire angular phase space, and thus this term results in soft-collinear double logarithms and hard-collinear single logarithms from the incoming leg $(a)$, \footnote{The collinear singularity to the incoming leg $(a)$ is at  $\eta\to+\infty$.} as well as soft wide-angle single logarithms from the whole dipole $(aj)$. The second term is similar to the first one (contains no collinear pole to the outgoing jet) but the integration is restricted to the interior of the outgoing hard jet, meaning that it results purely in soft wide-angle single logarithms from the dipole $(aj)$. The last term has a pure collinear pole to the outgoing jet, but contains a step function that restricts the gluon to be outside of it. This terms will thus result in a soft wide-angle single logarithm. In what follows we show how to integrate each term separately.

\subsubsection{First term}

We start with the integration
\begin{align}
&\mathcal{R}_{\delta}^{(aj)-1}(\bar{b})=\mathcal{C}_{aj}\int\frac{\as(\kappa_t^2)}{\pi}\,\frac{\d\kappa_t}{\kappa_t}\,\d\eta\,\frac{\d\phi}{2\pi}\,
\Theta\left(\kappa_t\sqrt{\omega_{aj}^k}\,|\sin\phi|-p_t/\bar{b}\right)\Theta\left(p_t-\sqrt{w_{aj}^{k}}\,\kappa_t\right)\times\notag\\
&\times\left[\frac{1}{2}\,\frac{\exp(\eta-y)}{\cosh(\eta-y)-\cos(\phi-\pi)}-\frac{1}{(\eta-y)^2+(\phi-\pi)^2}\right]\Theta\left[-\eta+y+B_a+\ln\frac{Q_{aj}^2}{k_t\,p_t}\right],
\end{align}
where we restored the $\eta$ and $\phi$ variables here for convenience and included hard-collinear emissions to the incoming leg $(a)$ via the last step function.

This contribution has both double and single logarithms, and is free from collinear logarithms to the outgoing jet. The double logarithms originate from soft-collinear emissions to the incoming leg $(a)$ corresponding to $\eta\to+\infty$, where we can approximate $\omega_{aj}^k\to 1$ and thus replace $k_t \to \kappa_t$ in the step function $\Theta\left(-\eta+y+B_a+\ln[Q_{aj}^2/\kappa_t\,p_t]\right)$. Furthermore, for the single-logarithmic part, the coefficients that multiply $\kappa_t$ in the two step functions $\Theta\left(\kappa_t\sqrt{\omega_{aj}^k}\,|\sin\phi|-p_t/\bar{b}\right)$ (i.e. the coefficient $\sqrt{\omega_{aj}^k}\,|\sin\phi|$) and $\Theta\left(p_t-\sqrt{w_{aj}^{k}}\,\kappa_t\right)$ (i.e. the coefficient $\sqrt{\omega_{aj}^k}$) only contribute at sub-leading accuracy, and can thus be substituted for any constant. We thus set $\omega_{aj}^k\to 1$ in all the step functions, and additionally take the average contribution of $\phi$ in the step function (in the region $\eta\to+\infty$) exploiting $\int_{0}^{2\pi}\ln|\sin\phi|\,\d\phi/2\pi=-\ln2$ as we observed in the previous subsection. Hence we may write at single-logarithmic accuracy
\begin{align}
&\mathcal{R}_{\delta}^{(aj)-1}(\bar{b})=\mathcal{C}_{aj}\int\frac{\as(\kappa_t^2)}{\pi}\,\frac{\d\kappa_t}{\kappa_t}\,\d\eta\,\frac{\d\phi}{2\pi}\,\Theta\left(\frac{1}{2}\,\kappa_t-p_t/\bar{b}\right)
\Theta\left(p_t-\kappa_t\right)\times\notag\\
&\times\left[\frac{1}{2}\,\frac{\exp(\eta-y)}{\cosh(\eta-y)-\cos(\phi-\pi)}-\frac{1}{(\eta-y)^2+(\phi-\pi)^2}\right]\Theta\left[-\eta+y+B_a+\ln\frac{Q_{aj}^2}{\kappa_t\,p_t}\right].
\end{align}
We can then integrate over $\eta$ and $\phi$ up to single logarithmic accuracy obtaining
\begin{align}
&\int_{-\infty}^{B_a+y+\ln(Q_{aj}^2/\kappa_t\,p_t)}\d\eta\,\frac{\d\phi}{2\pi}\left[\frac{1}{2}\,\frac{\exp(\eta-y)}{\cosh(\eta-y)-\cos(\phi-\pi)}-\frac{1}{(\eta-y)^2+(\phi-\pi)^2}\right] =\notag\\
&=B_a+\ln\frac{Q_{aj}^2}{p_t^2}-\ln\frac{\kappa_t}{p_t}-\ln (2\pi)\,.
\end{align}
Notice that the integration over $\omega_{aj}^{k}$ and $1/(R^2\,r^2)$ separately diverge when $r\to 0$, but the overall integration is finite. One can avoid this divergence by simply placing a cutoff $\epsilon$ on $\eta-y$ around $0$, and setting $\epsilon\to0$ at the end. Now we perform the integration over $\kappa_t$, which is very similar to that performed in subsection \ref{sec:wps}, to obtain the following result
\begin{align}
\mathcal{R}_{\delta}^{(aj)-1}(\bar{b})&=\mathcal{C}_{aj}\int\frac{\as(\kappa_t^2)}{\pi}\,\frac{\d\kappa_t}{\kappa_t}\,\Theta\left(\frac{\kappa_t}{2}-\frac{p_t}{\bar{b}}\right)\Theta\left(p_t-\kappa_t\right)
\left(B_a+\ln\frac{Q_{aj}^2}{p_t^2}-\ln\frac{\kappa_t}{p_t}-\ln(2\pi)\right)\notag\\
&=-\frac{\mathcal{C}_{aj}}{2}\,L\,\frac{1}{2\pi\beta_0\,\Lambda}\left[2\Lambda+\ln(1-2\Lambda)\right]+\frac{\mathcal{C}_{aj}}{2}\,\frac{K}{4\pi^2\beta_0^2}\left[\ln(1-2\Lambda)+\frac{2\Lambda}{1-2\Lambda}\right]-\notag\\
&-\frac{\mathcal{C}_{aj}}{2}\,\frac{\beta_1}{2\pi\beta_0^3}\left[\frac{1}{2}\ln^2(1-2\Lambda)+\frac{\ln(1-2\Lambda)+2\Lambda}{1-2\Lambda}\right]-\frac{\mathcal{C}_{aj}}{2}\,\ln2\,\frac{1}{\pi\beta_0}\,
\frac{2\Lambda}{1-2\Lambda}-\notag\\
&-\mathcal{C}_{aj}\,\frac{1}{2\pi\beta_0}\left[\ln \frac{Q_{aj}^2}{p_t^2}+B_a-\ln(2\pi)\right]\ln(1-2\Lambda)\,.
\end{align}

\subsubsection{Second term}

The second term to integrate is
\begin{align}
\mathcal{R}_{\delta}^{(aj)-2}(\bar{b})&=\mathcal{C}_{aj}\int\frac{\as(\kappa_t^2)}{\pi}\,\frac{\d\kappa_t}{\kappa_t}\,R^2\,r\,\d r\,\frac{\d\theta}{2\pi} \left(\frac{1}{R^2r^2}-w_{aj}^k\right)\Theta_{\inn}(k)\,\Theta\left(\sqrt{w_{aj}^{k}}\kappa_{t}|\sin\phi|-p_t/\bar{b}\right)\notag\\
&\times\Theta\left(p_t/\sqrt{w_{aj}^{k}}-\kappa_t\right).
\end{align}
This term produces only single logarithms as explained earlier, so we integrate over $\kappa_t$ using the one-loop QCD beta function. The three step functions in the integrand reduce to
$$\Theta(1-r^2)\,\Theta(\kappa_t-p_t/\bar{b})\,\Theta(p_t-\kappa_t)\,.$$
Performing the integrations over $\kappa_t$, $r$ and $\theta$ we obtain the result
\begin{equation}
\mathcal{R}_{\delta}^{(aj)-2}(\bar{b})=\mathcal{C}_{aj}\,\frac{1}{2\pi\beta_0}\ln(1-2\Lambda)\left(\frac{1}{8}R^2+\frac{1}{576}\,R^4+\mathcal{O}(R^8)\right).
\end{equation}

\subsubsection{Third term}

The last term to integrate is
\begin{align}
\mathcal{R}_{\delta}^{(aj)-3}(\bar{b})&=\mathcal{C}_{aj}\int\frac{\as(\kappa_t^2)}{\pi}\,\frac{\d\kappa_t}{\kappa_t}\,R^2\,r\,\d r\,\frac{\d\theta}{2\pi}\,\frac{1}{R^2r^2}\, \Theta_{\out}(k)\,\Theta\left(\sqrt{w_{aj}^{k}}\,\kappa_{t}\,|\sin\phi|-p_t/\bar{b}\right)\times\notag\\
&\times\Theta\left(p_t/\sqrt{w_{aj}^{k}}-\kappa_t\right).
\end{align}
Performing the integration over $\kappa_t$, $r$ and $\theta$ at single logarithmic accuracy gives
\begin{align}
\mathcal{R}_{\delta}^{(aj)-3}(\bar{b})&=\mathcal{C}_{aj}\,\frac{1}{2\pi\beta_0}\ln\frac{R}{2\pi}\ln(1-2\Lambda)\,.
\end{align}

The overall contribution of the dipole $(aj)$ to the radiator is
\begin{align}
\mathcal{R}_{\delta}^{(aj)}(\bar{b})&=-\frac{\mathcal{C}_{aj}}{2}\,L\,\frac{1}{2\pi\beta_0\,\Lambda}\left[2\Lambda+\ln(1-2\Lambda)\right]+\frac{\mathcal{C}_{aj}}{2}\,\frac{K}{4\pi^2\beta_0^2}
\left[\ln(1-2\Lambda)+\frac{2\Lambda}{1-2\Lambda}\right]-\notag\\
&-\frac{\mathcal{C}_{aj}}{2}\,\frac{\beta_1}{2\pi\beta_0^3}\left[\frac{1}{2}\ln^2(1-2\Lambda)+\frac{\ln(1-2\Lambda)+2\Lambda}{1-2\Lambda}\right]-\frac{\mathcal{C}_{aj}}{2}\,\ln2\,\frac{1}{\pi\beta_0}\,\frac{2\Lambda}{1-2\Lambda}-\notag\\
&-\mathcal{C}_{aj}\frac{1}{2\pi\beta_0}\left[\ln \frac{Q_{aj}^2}{p_t^2}+B_a\right]\ln(1-2\Lambda)+\notag\\
&+\mathcal{C}_{aj}\frac{1}{2\pi\beta_0}\ln(1-2\Lambda)\left(\ln R+\frac{1}{8}R^2+\frac{1}{576}\,R^4+\mathcal{O}(R^8)\right)\,.
\end{align}

We finally note that the contribution of the dipole $(bj)$ to the radiator is identical to that of the dipole $(aj)$, with the substitution $a\to b$.

\subsection{Assembled expression for the radiator}

The final expression for the total radiator is given by
\begin{equation}
\mathcal{R}_{\delta}(\bar{b})=\mathcal{R}_{\delta}^{\mathrm{coll.}}(\bar{b})+\mathcal{R}_{\delta}^{\mathrm{wide}}(\bar{b})\,,
\end{equation}
where the contribution corresponding to collinear (soft or hard) emissions from the incoming legs ($a$ and $b$) is expressed as
\begin{equation}
\mathcal{R}_{\delta}^{\mathrm{coll.}}(\bar{b})=\left(\mathcal{C}_a+\mathcal{C}_b\right)\left[L\,g_1(\alpha_s L)+g_2(\alpha_sL)-\ln2\,g'(\alpha_sL)\right]+\left(\mathcal{C}_a\,B_a+\mathcal{C}_b\,B_b\right)t(\alpha_sL)\,,
\end{equation}
with
\begin{align}
g_1&=-\frac{1}{2\pi\beta_0\,\Lambda}\left[2\Lambda+\ln(1-2\Lambda)\right],\\
g_2&=\frac{K}{4\pi^2\beta_0^2}\left[\ln(1-2\Lambda)+\frac{2\Lambda}{1-2\Lambda}\right]-\frac{\beta_1}{2\pi\beta_0^3}\left[\frac{1}{2}\ln^2(1-2\Lambda)+\frac{\ln(1-2\Lambda)+2\Lambda}{1-2\Lambda}\right],\\
g'&=\frac{\partial g}{\partial L}=\frac{2}{\pi\beta_0}\,\frac{\Lambda}{1-2\Lambda}\,,\label{eq:gps}\\
t&=-\frac{1}{\pi\beta_0}\ln(1-2\Lambda)\,,
\end{align}
and $g=L\,g_1+g_2$. The functions $g_1$ and $g_2$ can easily be obtained from the principles of general final-state resummation of ref. \cite{Banfi:2004yd}. \footnote{See eq. (3.6), and the expressions (A.4) and (A.6), with $a=1$ corresponding to the power of $k_t$ in the definition of our observable.} Additionally the function resumming soft wide-angle emissions from all the three hard legs is
\begin{align}
\mathcal{R}_{\delta}^{\mathrm{wide}}(\bar{b})&=\frac{1}{2}\,t(\alpha_s\,L)\,\sum_{(\alpha\beta)}\mathcal{C}_{\alpha\beta}\left(\ln\frac{Q^2_{\alpha\beta}}{p^2_t}-h_{\alpha\beta}(R)\right),
\end{align}
where the sum extends over the three dipoles and the jet-radius--dependent functions are given by
\begin{align}
h_{ab}(R)&=\frac{R^2}{2}\,,\\
h_{aj}(R)=h_{bj}(R)&=\ln R+\frac{1}{8}\,R^2+\frac{1}{576}\,R^4+\mathcal{O}(R^8)\,.
\end{align}
In the above we define $\mathcal{C}_a=(\mathcal{C}_{ab}+\mathcal{C}_{aj})/2$ and $\mathcal{C}_b=(\mathcal{C}_{ab}+\mathcal{C}_{bj})/2$. It turns out that $\mathcal{C}_i=\CF$ for quark leg $i$ and $\mathcal{C}_i=\CA$ for gluon leg $i$.

Finally, the derivative of the radiator with respect to $\ln \bar{b}$ (up to NLL accuracy) is given by
\begin{equation}
\mathcal{R}_{\delta}'(\bar{b}) = \frac{\partial \mathcal{R}_{\delta}}{\partial \ln\bar{b}} = (\mathcal{C}_a+\mathcal{C}_b) g'(\alpha_s L)\,,
\end{equation}
where $g'$ is defined in eq. \eqref{eq:gps}.

\section{Fixed-order expansion} \label{sec:AppB}

In order to expand the resummed distribution (given \emph{formally} by eq. \eqref{eq:final2}) at fixed order, we first need to relate the PDFs of flavour $i$ evaluated at the scale $\mu_{\mathrm{f}}^2\,\Delta^2$ to the PDFs at the factorisation scale $\mu_{\mathrm{f}}^2$. The relevant expression can be obtained by solving the DGLAP evolution equation \cite{Gribov:1972ri} at leading order to obtain \footnote{See ref. \cite{Banfi:2004yd}, page 59 and 56 for further details, including expressions of leading-order splitting functions.}
\begin{equation}
f_i\left(x,\Delta^2\,\mu_{\mathrm{f}}^2\right)=f_i\left(x,\mu_{\mathrm{f}}^2\right)\left(1-\frac{1}{f_i\left(x,\mu_{\mathrm{f}}^2\right)}\,\frac{\alpha_s}{\pi}\ln\frac{1}{\Delta}\,\sum_j
\int_x^1 P_{ij}(\xi)\,f_j\left(x/\xi,\mu_{\mathrm{f}}^2\right)\frac{\d\xi}{\xi}\right),
\end{equation}
where $P_{ij}(\xi)$ are the corresponding leading-order Altarelli-Parisi splitting functions. For a fixed factorisation scale the expansion of the resummed formula \eqref{eq:final2} at fixed order is
\begin{align}
\sigma_1(\Delta)&=\sum_{\delta}\int\d\B_{\delta}\,\frac{\d\sigma_{0\delta}}{\d\B_{\delta}}\,\Xi_\B\,
\left(1+G_{11}\,\frac{\alpha_s}{\pi}\,\ln\frac{1}{\Delta}+G_{12}\,\frac{\alpha_s}{\pi}\,\ln^2\frac{1}{\Delta}\right),\label{eq:expansion}
\end{align}
where
\begin{align}
G_{11}&=\left(\C_a+\C_b\right)2\ln2-2\left(\C_a\,B_a+\C_b\,B_b\right)+\C_{ab}\,\frac{R^2}{2}+(\C_{aj}+\C_{bj})\left(\ln R+\frac{R^2}{8}+\frac{R^4}{576}\right)-\notag\\
&-2\left(\C_{ab}\ln\frac{Q_{ab}}{p_t}+\C_{aj} \ln\frac{Q_{aj}}{p_t}+\C_{bj}\ln\frac{Q_{bj}}{p_t}\right)-P(\mathcal{B})\,,\\
G_{12}&=-(\C_a+\C_b)\,,
\end{align}
with the process-dependent (on $x_a$ and $x_b$) term (sum over $j$ implied)
\begin{equation}
P(\mathcal{B})=\frac{1}{f_a(x_a,\mu_{\mathrm{f}}^2)}\int_{x_a}^1P_{aj}(\xi)\,f_j(x_a/\xi,\mu_{\mathrm{f}}^2)\,\frac{\d\xi}{\xi}
+\frac{1}{f_b(x_b,\mu_{\mathrm{f}}^2)}\int_{x_b}^1P_{bj}(\xi)\,f_j(x_b/\xi,\mu_{\mathrm{f}}^2)\,\frac{\d\xi}{\xi}\,.
\end{equation}

\acknowledgments

\begin{itemize}
\item This work is supported by PRFU research project B00L02UN050120190001. We thank the Algerian Ministry of Higher Education and Scientific Research and DGRSDT for financial support.
\item Some of the numerical calculations in this paper have been performed in the High-Performance-Computing cluster at the University of Batna 2 (UB2-HPC).
\end{itemize}

\bibliographystyle{JHEP}
\bibliography{Refs}

\end{document}